\documentclass[a4paper,11pt]{article}
\usepackage{jheppub} 
\usepackage[T1]{fontenc} 
\usepackage{slashed}
\usepackage{scalefnt}
\usepackage{dsfont}

\newcommand{\lmt}{L_{mt}}
\newcommand{\EulerGamma}{\gamma_\mathrm{E}}
\newcommand{\abbrev}[1]{{\scalefont{.9}#1}}
\newcommand{\gff}{\abbrev{GFF}}
\newcommand{\be}{\begin{equation}}
\newcommand{\ee}{\end{equation}}
\newcommand{\Slash}[1]{{\ooalign{\hfil/\hfil\crcr$#1$}}}

\newcommand{\lt}{\left}
\newcommand{\rt}{\right}
\newcommand{\del}{\partial}

\newcommand{\non}{\nonumber \\}

\newcommand{\mbar}{\overline{m}}
\newcommand{\MSbar}{\overline{\rm MS}}
\newcommand{\ep}{\epsilon}

\newcommand{\dd}{\mathrm{d}}
\newcommand{\rme}{\mathrm{e}}

\newcounter{notecount}

\title{A new approach to quark mass determination using the gradient flow
}
\author[a]{Hiromasa Takaura,}
\author[b]{Robert V. Harlander,}
\author[c,d]{Fabian Lange}
\affiliation[a]{Center for Gravitational Physics and Quantum Information, Yukawa Institute for Theoretical Physics, Kyoto University,
Kyoto 606-8502, Japan}
\affiliation[b]{Institute for Theoretical Particle Physics and Cosmology, RWTH Aachen University, Sommerfeldstra\ss{}e 16, D-52056 Aachen, Germany}
\affiliation[c]{Physik-Institut, Universit\"at Z\"urich, Winterthurerstrasse 190, 8057 Z\"urich, Switzerland}
\affiliation[d]{PSI Center for Neutron and Muon Sciences, Forschungsstrasse 111, 5232 Villigen PSI, Switzerland}

\emailAdd{hiromasa.takaura@yukawa.kyoto-u.ac.jp,
  robert.harlander@rwth-aachen.de, fabian.lange@physik.uzh.ch
}

\abstract{ We propose a new method to determine quark masses using ratios of the
vacuum-expectation values (\abbrev{VEV}s) of flowed quark bilinear operators. They can be expressed as functions of the flow time $t$
  and the $\MSbar$ quark mass $\mbar$, which can then be determined by
  matching with the corresponding lattice results.  Motivated by this, we
  evaluate these \abbrev{VEV}s perturbatively through next-to-leading order in
  the strong coupling. We provide the results as expansions in the limits of
  small and large $\mbar^2 t$. To this end, we develop a new expansion
  technique based on the Laplace transform.  Additionally, we present
  numerical results with the exact mass dependence over a wide range of
  $\mbar^2t$.  We discuss the expected perturbative precision for the mass
  determination based on our next-to-leading order perturbative calculations,
  and possible non-perturbative corrections.  }

\preprint{YITP-25-89, P3H-25-037, TTK-25-15, ZU-TH 42/25}

\begin{document}
\maketitle
\flushbottom


\section{Quark mass determination using the gradient flow}\label{}

The gradient-flow formalism (\gff)
\cite{Narayanan:2006rf,Luscher:2010iy,Luscher:2013cpa} has been playing a
crucial role in recent developments in lattice \abbrev{QCD}.  It allows for
efficient scale setting, provides a gauge-invariant renormalization scheme
which can be implemented on the lattice as well as in dimensional
regularization, and enables one to assume the symmetries of the continuum even
in lattice calculations.

In this paper, we propose a way to determine the quark masses with the \gff.
Quark masses are fundamental parameters of \abbrev{QCD}, and their precise
determination is essential in contexts such as Higgs and flavor physics.  Our
proposal is to consider the ratios of the vacuum expectation values
(\abbrev{VEV}s) of the bilinear operators
\begin{equation}\label{SandR}
  \begin{aligned}
    S(t) \equiv \langle \bar{\chi}(t,x) \chi(t,x) \rangle \qquad{} \text{and} \qquad R(t) \equiv
    \left\langle \bar{\chi}(t,x) \overleftrightarrow{\slashed{\mathcal{D}}} \chi(t,x)
    \right\rangle
  \end{aligned}
\end{equation}
where $\chi(t,x)$ and $\bar{\chi}(t,x)$ denote the flowed quark fields.
Matrix elements of $S(t)$ and $R(t)$ are individually ultraviolet
(\abbrev{UV}) divergent in general, even after renormalizing the \abbrev{QCD}
parameters. However, the remaining \abbrev{UV} divergences can be fully
absorbed into a flowed-quark wave function renormalization factor
\cite{Luscher:2013cpa}.  Thus, the divergences cancel when considering ratios
of such matrix elements~\cite{Makino:2014taa}.  Therefore, the continuum limit
in a lattice determination of these ratios exists.  On the other hand, one can
calculate these ratios in perturbation theory, where the result is expressed
as a function of the flow time $t$, the strong coupling $\alpha_{\mathrm{s}}$,
and the quark mass $\mbar$, renormalized in the modified minimal-subtraction
scheme ($\MSbar$).  By matching the perturbative and lattice results, one
should thus be able to determine the quark mass $\mbar$.

The determination of the quark masses has been the subject of lattice
calculations for many years. We refer to
ref.~\cite{FlavourLatticeAveragingGroupFLAG:2024oxs} for an overview of
available results.  The main methodologies behind these studies are the
regularization-independent momentum subtraction (\abbrev{RI-MOM})
scheme~\cite{Martinelli:1994ty} or its variant, the regularization-independent
symmetric momentum subtraction (\abbrev{RI-SMOM}) scheme~\cite{Aoki:2007xm,Sturm:2009kb},
the quark-current (or current-like) correlator method~\cite{HPQCD:2008kxl}, or
the minimal renormalon-subtracted (\abbrev{MRS})
scheme~\cite{Brambilla:2017hcq}.  Similar to the \abbrev{RI-(S)MOM} schemes,
our method is also based on quark bilinear operators. However, while in the
\abbrev{RI-(S)MOM} schemes three-point functions such as $\langle
\bar{\psi}(z_1) \bar{\psi} \psi (x) \psi(z_2) \rangle$ are calculated to study
the renormalization of the bilinear operators, our proposal is based on the
flowed one-point functions $S(t)$ and $R(t)$ of eq.~\eqref{SandR}. Moreover,
the quantities of interest in our approach are manifestly gauge invariant and
thus no gauge-variant operators are required for the renormalization or the
short-distance (small-flow-time) expansion, which allows for a systematic
discussion of non-perturbative corrections.

For the quark mass determination in this framework, one has to match lattice
results with perturbative results.  While $S(t)$ including its mass dependence
is known exactly at leading order in perturbation
theory~\cite{Luscher:2013cpa} and $R(t)$ can easily be obtained in the same
manner, higher-order results are only known in the limit of a small quark mass
$\mbar$.  $S(t)$ vanishes in the strictly massless limit; its linear mass
term, together with the massless limit of $R(t)$, was calculated at
$\mathcal{O}(\alpha_{\mathrm{s}})$ in ref.~\cite{Makino:2014taa}\footnote{The
result for $S(t)$ was first presented in the arXiv versions v1 and v2 of
ref.~\cite{Makino:2014taa}, whereas $R(t)$ was presented in later arXiv versions and the published version.} and at $\mathcal{O}(\alpha_{\mathrm{s}}^2)$ in
ref.~\cite{Artz:2019bpr}. In addition, the leading mass correction of
$\mathcal{O}(\mbar^2 t)$ is available through
$\mathcal{O}(\alpha_{\mathrm{s}}^2)$ from ref.~\cite{Lange:2021vqg}.

However, the size of the dimensionless parameter $\mbar^2 t$ can span a wide
range, particularly when one is interested in heavy (i.e., charm and bottom)
quarks.  This can be seen as follows. Assuming a lattice spacing $a$, let us
impose the conditions
\begin{equation}\label{eq::form}
  \begin{gathered}
    a^2 \ll 8 t \ll \Lambda_{\rm QCD}^{-2}
  \end{gathered}
\end{equation}
on the range of the flow time $t$, where $\Lambda_{\rm QCD}\sim
0.3\,\text{GeV}$ is the mass scale of \abbrev{QCD} derived from dimensional
transmutation.  The first condition is necessary for reasonable lattice
simulations of eq.~\eqref{SandR}, the second one is imposed by perturbativity.
Then, if $a^{-1}=4$~GeV is assumed as a reference,\footnote{
When one is interested in the bottom quark mass,
even finer lattice spacings seem to be required. We consider
a typical lattice spacing available today.}
$0.1 \ll 8 \mbar_c^2 t \ll
20$ in the charm-quark case and $1.0 \ll 8 \mbar_b^2 t \ll 200$ in the
bottom-quark case.  Therefore, the known results valid for $\mbar^2 t \ll 1$
are not always sufficient.

In this paper, we evaluate the mass dependence of $S(t)$ and $R(t)$ at
two-loop level, i.e., at $\mathcal{O}(\alpha_{\mathrm{s}})$ in perturbation
theory.  We provide the first several terms of the small- and large-$\mbar^2 t$
expansions for these quantities.\footnote{These are updated results of our
proceedings article~\cite{Takaura:2024ejs}, where the first few terms of the
expansions for $S(t)$ were presented.}  Additionally, we present the full mass
dependence over a wide range of $\mbar^2 t$, based on a numerical computation
using \texttt{ftint}~\cite{Harlander:2024vmn}.

To expand loop integrals in either $m^2 t$ or $1/(m^2 t)$, we develop a new
method based on the Laplace transform.  The standard technique to expand loop
integrals under a certain hierarchy is known as ``strategy of regions''
\cite{Beneke:1997zp}. Its application to the \gff\ in the small-flow-time
limit has been described in ref.~\cite{Harlander:2021esn}. The opposite limit
should be accessible as well with this method, following analogous
considerations as described in ref.~\cite{Beneke:2023wmt}, albeit in a
completely different context.  The strategy employed here, on the other hand,
is very symmetric in its application to small and large flow times.

The paper is organized as follows.  In sec.~\ref{sec:calc} we explain our
methods for the perturbative calculations.  Sec.~\ref{sec:results} presents
our main results and a study of the perturbative behavior.  We also provide an
estimate of the achievable precision for the mass extraction using our
next-to-leading order perturbative results.  In sec.~\ref{sec:NP}, we discuss
the size of non-perturbative corrections and a way to properly suppress them
for light quarks.  Sec.~\ref{sec:conclusions} is devoted to our conclusions.
Appendices \ref{sec:ints}--\ref{sec:C} collect some detailed formul\ae\ and
discussion.  Our full perturbative results are provided in
app.~\ref{app:results} and also in ancillary files.


\section{Calculational method}
\label{sec:calc}


In sec.~\ref{sec:theoframe} we briefly summarize the perturbative approach to
the \gff{} and then obtain the expressions for $S(t)$ and $R(t)$ in terms of
scalar two-loop integrals in sec.~\ref{sec:vevs}.  These loop integrals are
evaluated using two approaches.  On the one hand, we expand them for
$m^2 t \ll 1$ and $m^2t \gg 1$, leading to semi-analytic expressions in these
two limits.  The methodologies adopted here are explained in
sec.~\ref{sec:exp}.  We also solve differential equations to evaluate some
integrals, as explained in sec.~\ref{sec:diffeq}.  On the other hand, we
evaluate the integrals numerically while keeping their full mass dependence,
as explained in sec.~\ref{sec:numerical}.  In sec.~\ref{sec:sftx}, we discuss
how the first few orders in the expansions for small $m^2t$ can also be
extracted from known results based on the small-flow-time expansion of the
relevant bilinear operators.


\subsection{Theoretical framework}
\label{sec:theoframe}

The \abbrev{QCD} action, before gauge fixing, is defined by
\begin{equation}\label{eq:calc:gaze}
  \begin{aligned}
    S_{\rm QCD}= \int \dd^d x  \lt(\frac{1}{4 g_0^2} F^a_{\mu \nu}(x) F^a_{\mu \nu}(x)+ \bar{\psi}(x) (\Slash{D}+m)\psi \rt)
  \end{aligned}
\end{equation}
where $g_0$ denotes the bare coupling and
\begin{equation}\label{eq:calc:dopa}
  \begin{aligned}
F_{\mu \nu}(x)=\del_{\mu} A_{\nu}(x)-\del_{\nu} A_{\mu}(x)+[A_{\mu}(x), A_{\nu}(x)]
\qquad{} \text{and} \qquad
D_{\mu}=\del_{\mu}+A_{\mu} .
  \end{aligned}
\end{equation}
We write the gauge field as $A_{\mu}=A^a_{\mu} T^a$, where the normalization
of the color generators $T^a$ in the fundamental representation is given by
\begin{equation}\label{eq:calc:babi}
  \begin{aligned}
{\rm Tr} \, (T^a T^b)=-\frac{1}{2} \delta^{ab} .
  \end{aligned}
\end{equation}
and
\begin{equation}\label{eq:calc:joel}
  \begin{aligned}
(T^a T^a)_{ij}=- C_{\mathrm{F}} \delta^{ij} .
  \end{aligned}
\end{equation}
with $C_{\mathrm{F}}=(N_{\mathrm{c}}^2-1)/(2 N_{\mathrm{c}})$.

In the \gff, one defines generalizations of the gauge and the fermion fields in
terms of the flow
equations~\cite{Narayanan:2006rf,Luscher:2010iy,Luscher:2013cpa}
\begin{align}
\partial_t \chi(t,x) &=(\mathcal{D}_{\mu} \mathcal{D}_{\mu}-\alpha_0
\partial_{\mu} B_{\mu}(t,x) )\chi(t,x) , \nonumber \\ \partial_t
\bar{\chi}(t,x) &=\bar{\chi}(t,x)\left(\overleftarrow{\mathcal{D}}_{\mu}
\overleftarrow{\mathcal{D}}_{\mu} +\alpha_0 \partial_{\mu} B_{\mu}(t,x) \right) ,
\nonumber \\ \partial_t B_{\mu}(t,x)&=\mathcal{D}_{\nu} G_{\nu
  \mu}(t,x)+\alpha_0 \mathcal{D}_{\mu} \partial_{\nu} B_{\nu}(t,x) ,
\label{eq:flow}
\end{align}
and the boundary conditions
\begin{equation}\label{eq:calc:febe}
  \begin{aligned}
    \chi(t=0,x)=\psi(x)\,,\qquad
    \bar{\chi}(t=0,x)=\bar{\psi}(x)\,,\qquad
    B_{\mu}(t=0,x)=A_{\mu}(x)\,.
  \end{aligned}
\end{equation}
The flowed covariant derivative is defined by
\begin{equation}\label{eq:calc:gowd}
  \begin{aligned}
    \mathcal{D}_{\mu}=\partial_{\mu}+B_{\mu}\,,\qquad
    \overleftarrow{\mathcal{D}}_{\mu}=
    \overleftarrow{\del}_{\mu}-B_{\mu}\,,\qquad
        \mathcal{D}_{\mu}=\partial_{\mu}+[B_{\mu}, \cdot]
  \end{aligned}
\end{equation}
for the flowed quark, anti-quark, and gauge field,
respectively.
Also, for flowed quark bilinear operators, we use
\be
\overleftrightarrow{\mathcal{D}}_{\mu}=\mathcal{D}_{\mu}-\overleftarrow{\mathcal{D}}_{\mu} .
\ee$
\alpha_0$ is a gauge fixing parameter
which cancels in physical quantities.  Throughout our calculation, we fix
$\alpha_0=1$ and employ Feynman gauge for the gluon propagator.


\subsection{Expressing the VEVs in terms of scalar integrals}
\label{sec:vevs}

Straightforward algebraic operations based on the perturbative method for
solving the flow equations~(\ref{eq:flow}) \cite{Luscher:2011bx} allow us to
express $S(t)$ and $R(t)$ at \abbrev{NLO} as
\begin{equation}
  \begin{aligned}
S|_{\rm
  2-loop} &=g_0^2 N_{\mathrm{c}} C_{\mathrm{F}} \Big[ 8 (d-2) I_1 -32 I_2 -16 I_3 -16 I_4\\&
  \qquad
  -8 I_5 +32
  I_7 -4(d-2) I_{10} +16 I_{11} \Big]\,,\\
R|_{\rm 2-loop}
&=g_0^2 N_{\mathrm{c}} C_{\mathrm{F}} \Big[ 16 J_1+(4d-8) J_2+2(8-4d) J_3+(16d-32) J_5\\
  &\qquad-64 J_6+32 J_7
  -32 J_8+16 J_9-32 J_{10}+64 J_{11}+32 J_{12}
  \\&\qquad-64 J_{13}+16 J_{14}
  +(8-4d) J_{15}
  -(8-4d) J_{16}-(8-4d) J_{17}
  \\&\qquad-(16-8d) J_{18}
  +(16-8d) J_{19}-32 J_{20} \Big]\,. \label{R2loop}
  \end{aligned}
\end{equation}
The two-loop scalar integrals $I_n$ and $J_n$ are listed in
app.~\ref{sec:ints}, where we use the short-hand notation
\begin{equation}\label{eq:calc:isai}
  \begin{aligned}
    \int_p \equiv \int \frac{\dd^d p}{(2 \pi)^d} \quad{} \text{and} \quad
    \int_{p,k} \equiv \int \frac{\dd^d p \, \dd^d k}{(2 \pi)^d (2 \pi)^d}\,,
  \end{aligned}
\end{equation}
with $d=4-2\epsilon$ the number of spacetime dimensions.
We selectively made use of integration-by-parts identities to express some integrals in terms of others~\cite{Tkachov:1981wb,Chetyrkin:1981qh,Artz:2019bpr}.


\subsection{Expansion by Laplace transform}
\label{sec:exp}


In this section, we explain how to expand the scalar integrals listed in
app.~\ref{sec:ints} in the limits of small and large $m^2 t$.  Our main
technique, described in sec.~\ref{sec:laplace}, is applied to the integrals $I_1$,
$I_2$, $I_3$, $I_4$, $I_5$, $I_6$, $I_8$, $I_{10}$, $J_1$, $J_2$, $J_3$, and
$J_{17}$.  The remaining integrals can either be expressed in terms of the
already evaluated ones, up to the addition of $m$-independent integrals that
can be directly evaluated, or are linked through differential equations
(\abbrev{DE}s), whose solutions are discussed in Sec.~\ref{sec:diffeq}.


\subsubsection{Laplace transform}
\label{sec:laplace}

It is convenient to write a
general integral depending on $m^2$ and $t$ as
\begin{equation}\label{eq:calc:hush}
  \begin{aligned}
    I(m^2,t) &= t^{-\alpha/2}\hat{I}(m^2t)\,,
  \end{aligned}
\end{equation}
where $\alpha$ is the mass dimension of the original integral.  We now perform
a Laplace transformation of the dimensionless integral $\hat{I}(z)$ with
respect to $z\equiv m^2t$:
\begin{equation}
  \begin{aligned}
    \tilde{I}(v) \equiv \int_0^{\infty} \mathrm{d}z\, z^{-v-1}
    \hat{I}(z)\,.\label{eq:Laplace}
  \end{aligned}
\end{equation}
The inverse transformation is given by
\begin{equation}
  \begin{aligned}
    \hat{I}(z)=\frac{1}{2 \pi \mathrm{i}} \int_{-\mathrm{i} \infty+v_0}^{\mathrm{i}
   \infty+v_0} \mathrm{d} v\, \tilde{I}(v)\,z^v , \label{invLaplace}
  \end{aligned}
\end{equation}
where $v_0$ is a real number chosen such that the integral of eq.~\eqref{eq:Laplace}
converges for $v=v_0$.  If $\hat{I}=\mathcal{O}(z^{a_s})$ for small $z$ and
$I=\mathcal{O}(z^{-a_l})$ for large $z$, one may choose $-a_l<v_0<a_s$.  This
also implies that $\tilde{I}(v)$ develops singularities in the $v$-plane which
encode the structure of the expansions of $I(z)$ in $z$ and in $1/z$, as we
will now explain.

For $z=m^2 t \ll 1$, we consider closing the integration contour of the inverse transform~\eqref{invLaplace}
in the right half of the $v$-plane, since $z^v\to 0$ as $v\to \infty$ in this case.
With the help of Cauchy's residue theorem we then arrive at
\begin{equation}
  \begin{aligned}
    \hat{I}(z\ll 1)&=-\sum_{v_{\rm sing}>v_0} {\rm Res} [\tilde{I}(v)
      z^v]|_{v=v_{\rm sing}} , \label{smallmassexpformula}
  \end{aligned}
\end{equation}
where we sum over the singularities of $\tilde{I}(v)$ located at
$v_{\rm sing}>v_0$.  ${\rm Res}$ represents the residue,
and the singularities are assumed to be poles.  In turn, for $z=m^2t \gg 1$, we close the
contour on the opposite side and obtain
\begin{equation}
  \begin{aligned}
    \hat{I}(z\gg 1)&=\sum_{v_{\rm sing}<v_0} {\rm Res} [\tilde{I}(v)
      z^v]|_{v=v_{\rm sing}} , \label{largemassexpformula}
  \end{aligned}
\end{equation}
where we sum over the singularities of $\tilde{I}(v)$ located at
$v_{\rm sing}<v_0$. The factor $z^v$ in eqs.~\eqref{smallmassexpformula} and
\eqref{largemassexpformula} allows us to distinguish individual orders of
the expansions of $\hat{I}(z)$ in $z$ and $1/z$, as we will see explicitly in
the following examples.  To obtain these expansions, we need to calculate the
singularities of the integrand in eq.~\eqref{invLaplace}.  Note that both the
small- and large-$m^2 t$ expansion can be obtained from the same quantity
$\tilde{I}(v)$.

Our approach is analogous to the ideas found in
refs.~\cite{Neubert:1994vb,Kitano:2022gzy}, for instance.  However, this is
the first application of the inverse Laplace transform to higher loop
integrals, employed for expanding them under certain hierarchy.
Eqs.~\eqref{eq:Laplace} and \eqref{invLaplace} are familiar from the standard
Mellin-Barnes method for Feynman integrals~\cite{Smirnov:1999gc,Tausk:1999vh}.
However, in this method transformations like eq.~\eqref{eq:Laplace} are
applied to rewrite every propagator by introducing auxiliary integration variables, whereas we introduce the single variable $v$ by integrating over the kinematic scale $z$.
Expansions in the context of standard Mellin-Barnes integrals have been discussed in refs.~\cite{Mishima:2018olh,Zhang:2024fcu}.


\subsubsection{One-loop example}\label{sec:oneloop}

Let us illustrate this method with a simple example.  At the leading order in
perturbation theory, $S$ is given by
\begin{equation}
  \begin{aligned}
    S|_{\rm 1-loop}=- 4 N_{\mathrm{c}} \int_p
    \frac{m}{m^2+p^2} \rme^{-2tp^2} \equiv -4N_{\mathrm{c}}\,t^{1/2-d/2}\,\hat{I}(m^2t)\,.
    \label{S1loop}
  \end{aligned}
\end{equation}
Applying the Laplace transform to the loop integrand yields a massless propagator with a power depending on $v$ and we can trivially perform the loop integral to obtain
\begin{align}
\tilde{I}(v)
&= t^{d/2}\int_p \rme^{-2 t p^2}\int_0^{\infty}
\mathrm{d} z\, \frac{z^{-v-1/2}}{z+tp^2}
= t^{d/2}\frac{\pi}{\cos(\pi v)} \int_p \lt( \frac{1}{tp^2} \rt)^{v+\frac{1}{2}} \rme^{-2t p^2} \non
&= \frac{\pi}{\cos(\pi v)} \frac{1}{(4\pi)^{2-\epsilon}} \frac{\Gamma(3/2-\epsilon-v)}{\Gamma(2-\epsilon)} 2^{v-3/2+\epsilon} .
\end{align}
In this example, one may take $v_0=0$, because each step of the
calculation is well defined.

Now we study the singularities of $\tilde{I}$.  There are two origins of
singularities.  One is $\pi/\cos(\pi v)$, a factor of the Laplace transform of
the loop {\it integrand}.  The other is $\Gamma(3/2-\ep-v)$, which emerges
after the loop integration.  The singularities of the latter reflect infrared or \abbrev{UV}
divergences of the integral over $p$, given in the second line of the
equation.  In this case, we do not have \abbrev{UV} divergences due to the factor
$\rme^{-2tp^2}$.

The positive ($v>v_0=0$) singularities of $\tilde{I}(v,t)$ contribute to
the small-$(m^2 t)$ expansion.
They are located at
\begin{equation}\label{eq:calc:hare}
  \begin{aligned}
    v = n-\frac{1}{2} \qquad
    \text{and}\qquad v=n+\frac{1}{2}-\epsilon\,,\qquad
    n\in\mathbb{N} = \{1,2,3,\ldots\}\,,
  \end{aligned}
\end{equation}
which give
\begin{align}
  -{\rm Res} [\tilde{I}(v) z^v]=
  \left\{\begin{array}{ll}
  \displaystyle \frac{z^{1/2}}{32
  \pi^2}
  +\mathcal{O}(\epsilon)\,, & v = 1/2\\[1em]
  \displaystyle \frac{z^{3/2}}{16\pi^2}
  \lt(\frac{1}{\epsilon}+1+3\log{2}+\log{\pi} \rt)+\mathcal{O}(\epsilon)\,,
  & v=3/2\\[1em]
  \displaystyle -\frac{z^{3/2-\ep}}{16 \pi^2}
 \lt(\frac{1}{\epsilon}+1-\EulerGamma+\log({4\pi})
 \rt)+\mathcal{O}(\epsilon)\,,& v = 3/2-\ep\\[1em]
 \qquad\vdots&
  \end{array}
  \right. .
\end{align}
The $1/\ep$ terms cancel between the $v=k$ and $v=k-\ep$ contributions.  As a
result, we obtain the small-$m^2 t$ expansion of eq.~\eqref{S1loop} as
\begin{align}
S|_{\rm 1-loop}
&=-\frac{N_{\mathrm{c}}}{8 \pi^2} \frac{m}{t}
\big[ 1+2 m^2 t\, \lmt  +4 (m^2 t)^2 (-1+\lmt)
+\mathcal{O}((m^2t)^3) \big]+\mathcal{O}(\ep)\,,
\end{align}
where
\begin{equation}\label{eq:calc:hugo}
  \begin{aligned}
    \lmt &= \log(2m^2t) + \EulerGamma\,.
  \end{aligned}
\end{equation}
The large-$m^2 t$ expansion arises from the negative singularities located at
\begin{equation}\label{eq:calc:ahom}
  \begin{aligned}
    v=-n+\frac{1}{2}\,,\qquad n\in\mathbb{N}\,.
  \end{aligned}
\end{equation}
In this case there are no $\ep$-shifted singularities.
We obtain the large-$m^2 t$ expansion
\begin{align}
S|_{\rm 1-loop}
&=-\frac{N_{\mathrm{c}}}{16 \pi^2} \frac{1}{m t^2}
\lt[ 1-\frac{1}{m^2 t}+\frac{3}{2} \frac{1}{(m^2 t)^2}
+\mathcal{O}((m^2t)^{-3}) \rt]+\mathcal{O}(\ep) .
\end{align}

These expansions agree with the result of ref.~\cite{Harlander:2021esn},
obtained with the strategy-of-regions method~\cite{Beneke:1997zp}.
In particular, the cancellation of
the $1/\ep$ terms has been
discussed within that context.
Since the exact result of eq.~\eqref{S1loop} can easily be obtained and reads
\be
S|_{\rm 1-loop}=-\frac{1}{(4 \pi)^{d/2}}4 N_{\mathrm{c}} m^{d-1}  \rme^{2 m^2 t} \Gamma(1-d/2, 2 m^2 t) ,
\ee
expanding it also allows us to verify the correctness of the new method.


\subsubsection{Two-loop examples}

Let us consider a few two-loop examples for the application of our
method. Like in the one-loop example of sec.~\ref{sec:oneloop}, we perform the
Laplace transform on the \textit{integrand} of the two-loop integrals, and
subsequently evaluate the loop integrations.
 It turns out that we can set
$v_0=0^-$ (slightly smaller than 0) in the inverse Laplace transform of eq.~\eqref{invLaplace} for all integrals.
We discuss a subtle point on this for $J_{17}$ at the end of this section.

\paragraph{The integral $I_2$.}
Our first example is the integral
\begin{equation}
  I_2=\int_0^t \dd s \int_0^s \dd s' \int_{p,k} \frac{m p^2}{k^2 (p^2+m^2)} \rme^{-(2t-s+s') p^2-(s+s') k^2-(s-s') (k-p)^2} .
\end{equation}
Using the formul\ae\ for the Laplace transform and the loop integrals of
app.~\ref{sec:C}, we find
\begin{align}
\tilde{I}_2(v)
&=-\frac{2^{-17/2-v+4 \epsilon} \pi^{-3+2 \ep} }{(-1+\ep) \Gamma(2-\ep)}
\frac{\Gamma(\frac{5}{2}-v-\ep)}{\cos(\pi v)} \non
&\quad{}
\times
\int_0^1 \int_0^1 \dd u\, \dd u'\, u^{\ep} (4-u(1-u')^2)^{-\frac{3}{2}+v+\ep}
{}_2 F_1 (1,v-\frac{1}{2}; 2-\ep, \frac{1}{4} u (1-u')^2),
\end{align}
where we introduced the dimensionless flow-time integration variables $s=tu$
and $s'=su'$, and $_2F_1$ is the hypergeometric function. Since the integral
over the flow-time parameters is finite, all singularities of $\tilde{I}_2(v)$
in $v$ are given by the factor $\Gamma(\frac{5}{2}-v-\ep)/\cos(\pi v)$, which
has only simple poles at
\begin{equation}\label{eq:calc:baal}
  \begin{aligned}
    v &= \pm \left(n-\frac{1}{2}\right) \qquad\text{and}\qquad
    v = n+\frac{3}{2}-\ep\,,\qquad n\in \mathbb{N}\,.
  \end{aligned}
\end{equation}
The residues of $\tilde{I}_2$ can thus be obtained by evaluating the $u,
u'$-integrals at these singular points. We use \texttt{HypExp}
\cite{Huber:2005yg,Huber:2007dx} to expand the hypergeometric functions in
$\ep$, which allows us to obtain the coefficients for the $1/\ep$ terms
analytically. It turns out that they cancel in the final result of
$I_2$. Concerning the constant term of $\mathcal{O}(\ep^0)$, we managed to
obtain an analytical (numerical) result for the residues of the negative
(positive) singularities.

\paragraph{The integral $I_5$.}
Our second example is
\begin{equation}
  I_5=\int_{p,k} \frac{m}{k^2(p^2+m^2)((p-k)^2+m^2)}\rme^{-tp^2-tk^2-t(k-p)^2} .
\end{equation}
As opposed to the integral $I_2$ discussed
above, the integral $I_5$ contains two massive propagators.
We thus first combine them using
Feynman parametrization:
\begin{equation}\label{eq:calc:akee}
  \begin{aligned}
    \frac{1}{(p^2+m^2)((p-k)^2+m^2)}=\int_0^1 \dd x \frac{1}{[x
        p^2+(1-x)(k-p)^2+m^2]^2} .
  \end{aligned}
\end{equation}
After the Laplace transform and the loop integrations, the Feynman parameter
integral can be evaluated analytically using \texttt{Mathematica}~\cite{Mathematica}. The only
remaining integration is then over the parameter $\beta$ (see
eq.~\eqref{formulatwoFeynman}) which we map to the interval $y\in[0,1]$ via
$\beta=y/(1-y)$. The $y$-integrand $f(y)$ is rather complicated and not shown
here. For positive $v$, it may develop singularities at $y=1$ which one can
write as
\begin{equation}\label{eq:calc:ikan}
  \begin{aligned}
    f(y) = \sum_{a=1}^2\frac{ (1-y)^{-v-a\ep +
        1/2}g_a(y)}{\Gamma(\frac{1}{2}+v)\cos(\pi
      v)}\,,
  \end{aligned}
\end{equation}
where the $g_a(y)$ are regular at $y=1$. To identify the singularities of the Laplace transform, we perform the $y$-integral as
\begin{equation}\label{eq:calc:cove}
  \begin{aligned}
    \int_0^1 \dd y\,f(y) &=
    \frac{1}{\Gamma(\frac{1}{2}+v)\cos(\pi v)}
    \sum_{a=1}^2\bigg[
    \int_0^1 \dd y\,(1-y)^{-v-a\ep + 1/2}\left[
      g_a(y) - \mathcal{T}_{1,N} g_a(y)\right]\\
    &\quad+ \sum_{k=0}^N \frac{(-1)^k}{k!}\frac{g_a^{(k)}(1)}{v-k-3/2+a\ep}\bigg]\,,
  \end{aligned}
\end{equation}
where $\mathcal{T}_{1,N}g_a(y)$ denotes the Taylor expansion of $g_a(y)$
around $y=1$ up to power $(1-y)^N$, $g_a^{(k)}(y)$ is the $k^\text{th}$
derivative of $g_a(y)$, and $N$ an integer part of $v-3/2$ (or $v-3/2\leq N<v-1/2$).

The remaining $y$ integral is finite and can be calculated numerically. It
contributes to the residues of the simple poles arising from the pre-factor
$1/(\Gamma(1/2+v) \cos{(\pi v)})$ which are
located at $v=n/2$ with $n\in\mathbb{N}$. On the other hand, one can directly
read off the residues $(-1)^k g_a^{(k)}(1)/k!$ of the poles at $v=k+3/2-a\ep$ arising
from the second term in square brackets.

Similarly, for negative $v$, the integrand $f(y)$ may develop singularities at
$y=0$, which can be treated in an analogous manner. Specifically, we find
\begin{equation}\label{eq:calc:cove1}
  \begin{aligned}
    \int_0^1 \dd y\,f(y) &=
    \frac{1}{\Gamma(\frac{1}{2}+v)\cos(\pi v)}
    \bigg[
    \int_0^1 \dd y\,y^{v+1/2}\left[
      h(y) - \mathcal{T}_{0,N} h(y)\right]\\
    &\quad+ \sum_{k=0}^N \frac{1}{k!}\frac{h^{(k)}(0)}{v+3/2+k}\bigg]\,,
  \end{aligned}
\end{equation}
with the integer $N=-v-3/2$.
Note that, since
the pre-factor only develops poles at positive $v$, the residues at negative
$v$ are given entirely by the Taylor expansion coefficients $h^{(k)}(0)$ of
the function $h(y)$ which can be easily evaluated analytically.

We managed to solve all the integrals $I_n$ and $J_n$ targeted by this method.
For $I_7$, $I_{11}$,
$J_{13}$, and $J_{15}$, we employ differential equations, as will be
explained in sec.~\ref{sec:diffeq}.

\paragraph{The integral $J_{17}$.} This integral involves an additional
complication, in that its Laplace transform contains singularities at $v=0$
and $v=-\ep$, meaning that $v_0$ in eq.~\eqref{invLaplace} has to be chosen carefully.
As argued in app.~\ref{sec:J17}, the choice $-\ep<v_0<0$ properly separates the small- and large-$(m^2 t)$ expansions, assuming $\ep>0$.

\paragraph{Spurious $1/t^2$ terms.} Some of the $J_n$ develop terms $\sim 1/t^2$ in the large-$m^2t$ expansion, even though the
one-loop result of $R(t)$ starts at order $1/t^3$ only:
\begin{equation}\label{eq:calc:jefe}
  \begin{aligned}
    R|_{\rm 1-loop}=-\frac{N_{\mathrm{c}}}{8 \pi^2} \frac{1}{m^2 t^3}
    \lt(1-\frac{3}{2 m^2t}+\frac{3}{(m^2 t)^2}+\mathcal{O}((m^2t)^{-3})
    \rt)+\mathcal{O}(\ep) ,
  \end{aligned}
\end{equation}
It turns out though that all $1/t^2$-terms indeed cancel among the
$J_n$ when combining them to $R(t)$ through eq.~\eqref{R2loop}.



\subsection{Differential equations}
\label{sec:diffeq}

To calculate the remaining integrals $I_7$, $I_{11}$, and $J_{13}$, it is more
convenient to make use of differential equations (\abbrev{DE}s) rather than
applying the Laplace transform.  In particular, $I_7$ has a flow-time integral
in addition to two massive propagators, which makes the calculation using the
Laplace transform more complicated.  It is easy to see that
\begin{equation}
  \begin{aligned}
    (\del_t-2m^2) I_7&=m^2
    (I_5-2I_6)\,,\\ (\del_t-2m^2) I_{11}&=-2 m^2 I_8\,,
    \label{DEI11}
  \end{aligned}
\end{equation}
$I_6$ is introduced for this purpose, although it does not directly contribute
to $S(t)$.

Let us consider a general \abbrev{DE} of the form
\begin{equation}
  \begin{aligned}
    (\del_t-2m^2) I(t)=I_{\rm ext}(t) ,
  \end{aligned}
\end{equation}
where the expansions of $I_{\rm ext}(t)$ are assumed to be known.  By
introducing $i(t)$ as
\begin{equation}\label{eq:calc:elik}
  \begin{aligned}
    I(t) \equiv \rme^{2 m^2 t} i(t) ,
  \end{aligned}
\end{equation}
we can derive the \abbrev{DE}
\begin{equation}
  \begin{aligned}
    i'(t)=\rme^{-2 m^2 t} I_{\rm ext}(t) \label{DEfori}
  \end{aligned}
\end{equation}
for $i(t)$.

We obtain the small-$m^2 t$ expansion of $I$ as follows.  Using the known
result of the small-$m^2 t$ expansion of $I_{\rm ext}(t)$, eq.~\eqref{DEfori}
is given in the form
\begin{equation}
  \begin{aligned}
    i'(t)= \sum_{k=0}^{\infty} (a_k m^{2k+3} t^{-1+k+2
      \epsilon}+b_k m^{2 (k-\epsilon)+3} t^{-1+k+\epsilon}+c_k m^{2 (k-2
      \epsilon)+3}t^{-1+k} ) . \label{iprime}
  \end{aligned}
\end{equation}
Note that we keep the powers in $t$ exact without expanding them in $\ep$ at
this stage, while the coefficients $a_k$, $b_k$ and $c_k$ are expanded in
$\ep$.  Solving this \abbrev{DE}, we obtain
\begin{equation}
  \begin{aligned}
    i(t)=\sum_{k=0}^{\infty}
    \left(\frac{a_k}{k+2 \epsilon} m^{2k+3} t^{k+2
      \epsilon}+\frac{b_k}{k+\epsilon} m^{2 (k-\epsilon)+3}
    t^{k+\epsilon}+\frac{c_k}{k} m^{2 (k-2 \epsilon)+3} t^k
    \right)+\text{const.} \label{solutioni}
  \end{aligned}
\end{equation}
We note that the coefficient $a_0$ is required up to $\mathcal{O}(\ep)$ in
order to obtain the result of $I(t)=\rme^{2 m^2 t} i(t)$ up to
$\mathcal{O}(\ep^0)$.  This means that the integrals on the right-hand side of
eqs.~\eqref{DEI11} need to be evaluated through $\mathcal{O}(\ep)$ using the
Laplace transform method described in sec.~\ref{sec:laplace}.  We also note
that our calculation gives $b_0=c_0=0$ for $I_7$ and $I_{11}$.

The constant term in eq.~\eqref{solutioni} is determined by
\begin{equation}\label{eq:calc:gimp}
  \begin{aligned}
    \text{const.}=i(0)=I(t=0)\,,
  \end{aligned}
\end{equation}
with an explicit calculation of $I(t=0)$, which corresponds to a regular
two-loop massive tadpole integral which can be calculated by standard methods.

Concerning the large-$m^2t$ expansion, we can write
\begin{align}
i(t)
&=-\int_{t}^{\infty} \dd t'\, i'(t')
=-\int_{t}^{\infty} \dd t'\, I_{\rm ext}(t') \rme^{-2 m^2t'} \non
&=-\rme^{-2 m^2t} \frac{I_{\rm ext}(t)}{2m^2}-\frac{1}{2 m^2}
\int_{t}^{\infty} \dd t' \, I'_{\rm ext}(t') \rme^{-2m^2t'}\label{eq:it}
\end{align}
where we used that $t^n \rme^{-2m^2t}\to 0$ for $t\to \infty$ and arbitrary
$n$. Furthermore, in the first step we assumed that $i(t)=\rme^{-2m^2t}I(t)\to 0$
for $t\to \infty$, which is certainly true as this limit does not introduce
any singularities in a flow-time integral $I(t)$.
Repeatedly iterating the
last step in eq.~\eqref{eq:it}, we arrive at
\begin{equation}
  \begin{aligned}
    I(t)=-\lt( \frac{I_{\rm ext}(t)}{2m^2}+ \frac{I'_{\rm ext}(t)}{(2m^2)^2}+ \frac{I''_{\rm ext}(t)}{(2m^2)^3}+\cdots \rt) . \label{largem2tDE}
  \end{aligned}
\end{equation}

Following this procedure, we obtain the small- and large-$m^2t$ expansions of
$I_7$ and $I_{11}$.
In an analogous manner, we use the \abbrev{DE}
\begin{equation}\label{eq:calc:kama}
  \begin{aligned}
    (\del_t-2m^2) J_{13}=m^2 (J_4-2 J_{12})\,,\qquad
  \end{aligned}
\end{equation}
to determine the expansions of $J_{13}$.

We remark that similar \abbrev{DE}s can be used to obtain or check
some of the results. For
example, $J_{15}$ and $J_{18}$ are related via
\begin{equation}\label{eq:calc:dime}
  \begin{aligned}
    (\del_t-2m^2) J_{18} = - 2m^2 J_{15}\,.
  \end{aligned}
\end{equation}


\subsection{Numerical evaluation}
\label{sec:numerical}

Since we expect the asymptotic expansions of the integrals to fail in the
intermediate region, we also evaluate the integrals numerically on a
one-dimensional grid in $m^2 t$. This is achieved with the help of the program
\texttt{ftint}~\cite{Harlander:2024vmn}. It uses Schwinger parametrization of
the momentum integrations in order to map the integrals onto a
multi-dimensional hypercube. The integrand becomes the product of polynomials
in the Schwinger and flow-time parameters, raised to $d$-dependent powers.
Such integrals can be evaluated algorithmically by sector
decomposition~\cite{Binoth:2000ps,Binoth:2003ak}, which expresses them in
terms of a Laurent series in $d-4$ whose coefficients are convergent integrals
which can be evaluated numerically. For the sector decomposition and the
numerical integration, \texttt{ftint} relies on
\texttt{pySecDec}~\cite{Borowka:2015mxa,Borowka:2017idc,
  Borowka:2018goh,Heinrich:2023til}.  Evaluation of all integrals to a
relative precision of $10^{-8}$ or better on a grid of 200\,points takes only
a few hours on a currently average computer.



\subsection{Small-flow-time expansion}
\label{sec:sftx}

In addition to a direct calculation of the mass expansion outlined above, one
can also determine the small-$m^2 t$ expansion order-by-order by employing the
small-flow-time expansion on the operator level~\cite{Luscher:2011bx} as was
briefly mentioned in ref.~\cite{Lange:2021vqg}.  This effectively amounts to
an operator product expansion with $t\to0$.  For $S(t)$ it reads
\begin{equation}
  \bar{\chi}(t,x) \chi(t,x) = \zeta_\text{S}^{(1)}(t) \frac{m}{t} \mathds{1} + \zeta_\text{S}^{(3)}(t) m^3 \mathds{1} + \zeta_\text{S}(t) \bar{\psi}(x) \psi(x) + \mathcal{O}(t)
  \label{eq:S-sftx}
\end{equation}
up to mass dimension three~\cite{Luscher:2013vga} where the small-flow-time expansion separates local, unflowed operators from flow-time dependent matching coefficients.
Eq.~\eqref{eq:S-sftx} holds both on the bare and the renormalized level.
By taking the \abbrev{VEV} we can obtain $S(t)$ in terms of the matching coefficients $\zeta(t)$ and \abbrev{VEV}s of unflowed operators.

Similarly, the small-flow-time expansion for $R(t)$ reads
\begin{equation}
  \begin{split}
    \hspace*{-0.15em}
    \bar{\chi}(t,x) \overleftrightarrow{\slashed{\mathcal{D}}} \chi(t,x) &= \zeta^{(0)}_{2}(t) \frac{1}{t^2} \mathds{1} + \zeta^{(2)}_{2}(t) \frac{m^2}{t} \mathds{1} \\
    &\phantom{=} + \zeta_{21}(t) \frac{1}{g^2} F^a_{\mu\nu} (x) F^a_{\mu\nu} (x) + \zeta_{22}(t) \bar{\psi}(x) \overleftrightarrow{\slashed{D}} \psi(x) + \zeta_{23}(t) m^4 \mathds{1} + \mathcal{O}(t)
  \end{split}
  \label{eq:R-sftx}
\end{equation}
up to mass dimension four~\cite{Harlander:2020duo}.
Using the equations of motion we can replace $\bar{\psi}(x) \overleftrightarrow{\slashed{D}} \psi(x) = - 2 m \bar{\psi}(x) \psi(x)$ so that the same operator contributes to both expansions.

$\zeta_\text{S}^{(1)}(t)$ in eq.~\eqref{eq:S-sftx} corresponds to the
leading contribution to $S(t)$ and was computed through \abbrev{NNLO} in
$\alpha_{\mathrm{s}}$ in refs.~\cite{Makino:2014taa,Artz:2019bpr}.\footnote{We again
refer to the arXiv version v2 of ref.~\cite{Makino:2014taa}.}
$\zeta_\text{S}^{(3)}(t)$ is available through \abbrev{NNLO} from
ref.~\cite{Borgulat:2023xml}, and $\zeta_\text{S}(t)$ through \abbrev{NNLO}
from
refs.~\cite{Hieda:2016lly,Mereghetti:2021nkt,Borgulat:2022cqe,Borgulat:2023xml}.
The \abbrev{VEV} of $\bar{\psi}(x) \psi(x)$ was computed through \abbrev{NNLO}
already a long time
ago~\cite{Broadhurst:1981jk,Spiridonov:1988md,Harlander:diss}, but only for a
theory with a single, massive quark flavor, i.e.\ without additional massless
quarks.

Similarly, $\zeta^{(0)}_{2}(t)$ of eq.~\eqref{eq:R-sftx} corresponds to the
leading contribution to $R(t)$ and is available through \abbrev{NNLO} from
refs.~\cite{Makino:2014taa,Artz:2019bpr}.  $\zeta^{(2)}_{2}(t)$ was computed through \abbrev{NNLO}
in ref.~\cite{Harlander:2020duo}.  In the same reference, $\zeta_{23}(t)$ was
computed to the same order.  $\zeta_{21}(t)$ and $\zeta_{22}(t)$ through
\abbrev{NNLO} are available from
refs.~\cite{Makino:2014taa,Harlander:2018zpi,Harlander:2020duo}.  Again, the \abbrev{VEV} of
$g^{-2} F^a_{\mu\nu} (x) F^a_{\mu\nu} (x)$ was calculated through \abbrev{NNLO} a
long time ago~\cite{Braaten:1991qm,Chetyrkin:1994ex,Harlander:diss}, but also
in a theory without additional massless quarks. Since the expansions of both
$\zeta_{21}(t)$ and the \abbrev{VEV} of $g^{-2} F^a_{\mu\nu} (x) F^a_{\mu\nu}
(x)$ start at $\mathcal{O}(\alpha_{\mathrm{s}})$, they only
contribute from \abbrev{NNLO} onwards and are required to lower orders
than the other contributions.

We use this strategy to extract analytic expressions up to
$\mathcal{O}(\alpha_{\mathrm{s}} m^3)$ for $S(t)$ and up to
$\mathcal{O}(\alpha_{\mathrm{s}} m^4)$ for $R(t)$, confirming our numerical
expansions to this order.  $C^{S, \ll 1}_{1,0}$, $C^{S, \ll 1}_{1,1}$, $C^{R,
  \ll 1}_{1,0}$, $C^{R, \ll 1}_{1,1}$, and $C^{R, \ll 1}_{1,2}$ in
app.~\ref{app:results} are obtained in this manner.

Extending the extraction to \abbrev{NNLO} in $\alpha_{\mathrm{s}}$ is more
subtle: Eqs.~\eqref{eq:S-sftx} and \eqref{eq:R-sftx} in this form implicitly
assume a sum over the quark flavors for all quark operators.  While up to
\abbrev{NLO} there is no mixing between heavy- and light-quark flavors and
this subtlety is irrelevant, it becomes important to distinguish the flavors
at \abbrev{NNLO}.  However, the published results of
refs.~\cite{Borgulat:2023xml,Harlander:2020duo} do not contain sufficient
information to fully disentangle the contributions.  In addition, the full
\abbrev{NNLO} result for $\langle \bar{\psi}(x) \psi(x) \rangle$ is missing as mentioned
above.\footnote{The required expressions could be easily obtained with current
methods, but their calculation is beyond the scope of this paper.}



\section{Results}
\label{sec:results}

\subsection{Perturbative results}
The renormalized coupling and mass in the $\MSbar$ scheme are given through
\begin{align}
g_0&=\lt( \frac{ \mu \rme^{\EulerGamma/2}}{\sqrt{4\pi}} \rt)^{\epsilon} Z_g  g(\mu^2) , \non
m&=Z_m \overline{m}(\mu) , \label{couplingrenom}
\end{align}
where
\begin{align}
Z_g&=1+\mathcal{O}(\alpha_{\mathrm{s}}) , \non
Z_m&=1-\frac{\alpha_{\mathrm{s}}(\mu^2)}{4 \pi} \frac{6 C_{\mathrm{F}}}{2 \epsilon} +\mathcal{O}(\alpha_{\mathrm{s}}^2) ,
\end{align}
with $\alpha_{\mathrm{s}}(\mu^2)=g^2(\mu^2)/(4 \pi)$.
In addition, the renormalized flowed quark fields in the $\MSbar$ scheme are defined by \cite{Luscher:2013cpa}
\be
\chi_{\mathrm{R}}(t,x)=Z^{1/2}_{\chi} \chi(t,x) \qquad{} \text{and} \qquad
\bar{\chi}_{\mathrm{R}}(t,x)=Z^{1/2}_{\chi} \bar{\chi}(t,x) , \label{waverenom}
\ee
where
\be
Z_{\chi}=1+\frac{\alpha_{\mathrm{s}}}{4 \pi} \frac{6 C_{\mathrm{F}}}{2 \epsilon}+\mathcal{O}(\alpha_{\mathrm{s}}^2) . \label{waverenomfac}
\ee
Unless otherwise stated, $\mbar$ denotes $\mbar(\mu)$ in what follows.

In this section, we analyze our two-loop results of the renormalized \abbrev{VEV}s
\begin{equation}\label{eq::cara}
  \begin{aligned}
    \overline{S}&:=\langle \bar{\chi}_{\mathrm{R}}(t,x) \chi_{\mathrm{R}}(t,x) \rangle\,,\\
    \overline{R}&:=\left\langle \bar{\chi}_{\mathrm{R}}(t,x) \overleftrightarrow{\slashed{\mathcal{D}}}
    \chi_{\mathrm{R}}(t,x) \right\rangle
  \end{aligned}
\end{equation}
in the $\MSbar$ scheme. The explicit expansions in the limits of large and
small $m^2t$ are presented in app.~\ref{app:results} and available in \texttt{Mathematica} readable format in ancillary files accompanying this manuscript.
We checked that the leading terms in the small-$\mbar^2 t$ limit are
correctly reproduced \cite{Makino:2014taa,Artz:2019bpr,Lange:2021vqg}.\footnote{We again
refer to the arXiv version v2 of ref.~\cite{Makino:2014taa} for $S(t)$.}
In addition, we verified that they
satisfy the renormalization group equations
\begin{align}
\mu^2 \frac{\dd}{\dd \mu^2} \langle \bar{\chi}_{\mathrm{R}}(t,x) \chi_{\mathrm{R}}(t,x) \rangle
&=\gamma_{\bar{\chi} \chi} \langle \bar{\chi}_{\mathrm{R}}(t,x) \chi_{\mathrm{R}}(t,x) \rangle ,
\non \mu^2 \frac{\dd}{\dd \mu^2} \left\langle \bar{\chi}_{\mathrm{R}}(t,x)
\overleftrightarrow{\slashed{D}} \chi_{\mathrm{R}}(t,x) \right\rangle &=\gamma_{\bar{\chi}
  \chi} \left\langle \bar{\chi}_{\mathrm{R}}(t,x) \overleftrightarrow{\slashed{D}}
\chi_{\mathrm{R}}(t,x) \right\rangle ,
\end{align}
where $\gamma_{\bar{\chi} \chi}=-3 C_{\mathrm{F}}
\alpha_{\mathrm{s}}/(4\pi)+\mathcal{O}(\alpha_{\mathrm{s}}^2)$, which follow from
eqs.~\eqref{waverenom} and \eqref{waverenomfac}.\footnote{Note also
that $\mu^2 \dd \alpha_{\mathrm{s}}/\dd\mu^2=-\ep \alpha_{\mathrm{s}}+\mathcal{O}(\alpha_{\mathrm{s}}^2)$, following from eq.~\eqref{couplingrenom}.}

\begin{figure}
\begin{minipage}{0.5\hsize}
\begin{center}
    \includegraphics[width=7.5cm]{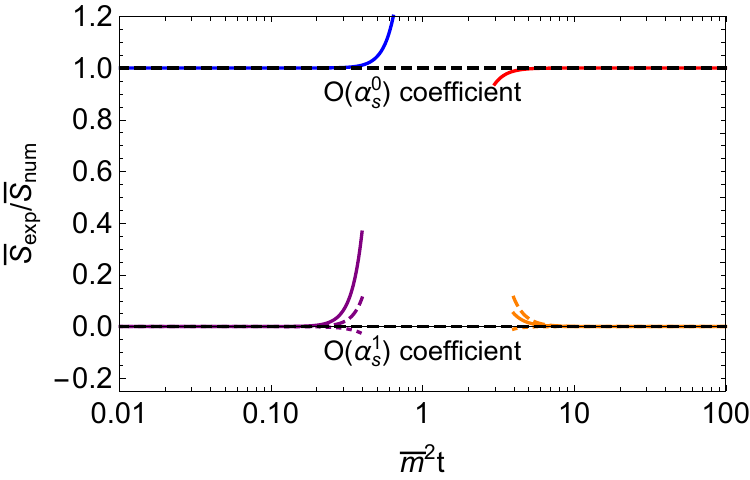}
\end{center}
\end{minipage}
\begin{minipage}{0.5\hsize}
\begin{center}
    \includegraphics[width=7.5cm]{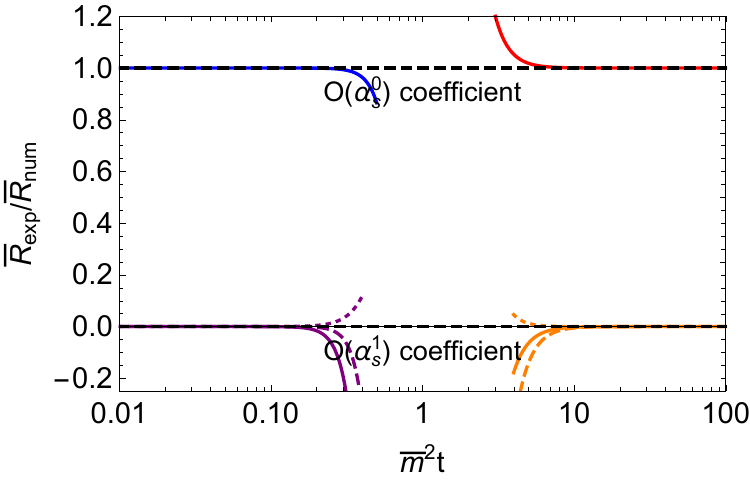}
\end{center}
\end{minipage}
\caption{The perturbative coefficients of the ratio of $\overline{S}$ (left)
and those of $\overline{R}$ (right), obtained using two approaches,
i.e., expansions and numerical evaluation, as a function
of $\mbar^2 t$, cf.~eq.~\eqref{eq:sexp}.
The lines in the region $\mbar^2 t <1$ correspond
to the small-$\mbar^2 t$ expansion (blue and purple lines)
and those in $\mbar^2 t >1$ to the large-$\mbar^2 t$ expansion (red and orange lines).
Since the $\mathcal{O}(\alpha_{\mathrm{s}})$ coefficient also depends on $\mu/\mbar$
in addition to $\mbar^2 t$, we show results with a few choices:
$\mu/\mbar=100$ (purple solid), $5$ (purple dashed), $1$ (purple dotted),
and $\mu/\mbar=1$ (orange solid), $2$ (orange dashed), $1/2$ (orange dotted).}
\label{fig:ExpvsNum}
\end{figure}

Another strong check of our results is the consistency of the expansions
with the numerical evaluation of the
integrals described in sec.~\ref{sec:numerical}, of course.  We provide these numerical
results for $\mbar^2 t \in [2^{-11}, 2^{11}]$ in ancillary
files as well. The explanation of the files is given in app.~\ref{app:results}.
The left part of fig.~\ref{fig:ExpvsNum} shows the perturbative coefficients of
$\overline{S}_{\rm exp}/\overline{S}_{\rm num}$, where $\overline{S}_{\rm
  exp}$ represents the small- and large-$\mbar^2 t$ expansion, whereas
$\overline{S}_{\rm num}$ is the numerical result retaining the full mass
dependence.  Schematically, the ratio is given by
\begin{align}
  \begin{split}
    \label{eq:sexp}
    \frac{\bar{S}_{\rm exp}}{\bar{S}_{\rm num}}
    &=\frac{\bar{S}_{\rm exp,LO}+\bar{S}_{\rm exp,NLO} \alpha_s+\mathcal{O}(\alpha_s^2)}{\bar{S}_{\rm num,LO}+\bar{S}_{\rm num,NLO} \alpha_s+\mathcal{O}(\alpha_s^2)}\\
    &=\frac{\bar{S}_{\rm exp,LO}}{\bar{S}_{\rm num,LO}}
    +\frac{\bar{S}_{\rm exp, NLO} \bar{S}_{\rm num, LO}-\bar{S}_{\rm exp, LO} \bar{S}_{\rm num, NLO}}{\bar{S}_{\rm num,LO}^2} \alpha_s+\mathcal{O}(\alpha_s^2) .
  \end{split}
\end{align}
We observe the expected behavior, where the $\mathcal{O}(\alpha_{\mathrm{s}}^0)$
coefficient approaches one and the $\mathcal{O}(\alpha_{\mathrm{s}})$ coefficient
approaches zero in the small- and large-$\mbar^2 t$ limits. Similarly, the
right figure shows the analogous result for $\overline{R}$.  The small-$\mbar^2
t$ expansion given here is valid for $\mbar^2 t \lesssim 0.2$, and the
large-$\mbar^2 t$ expansion is valid for $\mbar^2 t \gtrsim 6$.

Let us make a remark on the large-$\mbar^2 t$ expansion.
At one loop, the all-order series in $1/(m^2 t)$ reads \cite{Harlander:2021esn}
\be
S|_{\rm 1-loop}=-\frac{N_{\mathrm{c}}}{16 \pi^2} \frac{1}{\mbar t^2} \sum_{k=0}^{\infty} (-1)^k
\frac{(k+1)!}{(2 \mbar^2 t)^k} \,.
\ee
This is a sign-alternating divergent series, whose convergence radius is zero.
Therefore, extending the large-$\mbar^2 t$
expansion to higher orders does not necessarily improve its approximation.
The series gives the optimal approximation when it is truncated
at an optimal order $k_*$, which is estimated as
\be
k_* \sim 2 \mbar^2 t-2
\ee
depending on the size of $\mbar^2 t$ one is interested in.

\subsection{Lattice observables and uncertainty estimates for quark mass extractions}
\label{sec:pertres}

From the discussion at the beginning of sec.~\ref{sec:results} it is clear
that ratios of \abbrev{VEV}s of bilinear operators are finite and renormalization group invariant.
Specifically, we consider three observables
\begin{equation}\label{eq:resu:gulf}
  \begin{aligned}
    r_a(m) =\frac{S(t,m)}{R(t,m)}\,,\qquad
    r_b(m) = \frac{R(t,m)}{R(t,m=0)}\,,\qquad
    r_c(m) = m\frac{\dd}{\dd m}\left(\frac{S(t,m)}{R(t,m)}\right)
  \end{aligned}
\end{equation}
as examples.
All of them should be accessible on the lattice (the $m=0$ limit
may have to be approximated by a very small quark mass though). Their
perturbative expressions can be obtained from our expansions of $\bar{S}$ and
$\bar{R}$. They can also be obtained from the purely numerical results of
sec.~\ref{sec:numerical} ($r_c$ requires a numerical derivative which is
unproblematic though).

To see how these observables behave in the small- and large-$\mbar^2 t$
limits, we show the first few terms of the expansions in these limits.
We express the perturbative coefficients in terms of
\be
L_t \equiv \log(2 \mu^2 t)+\EulerGamma \qquad{} \text{and} \qquad
L_m \equiv \log(\mu^2/\mbar^2) .
\ee
For
$r_a$ the perturbative series reads
\begin{align}
r_a(\mbar)&=\mbar t \bigg\{ 1+C_{\mathrm{F}} \alpha_{\mathrm{s}}(\mu^2) (0.276671+0.238732
  L_t) \non &\qquad{}\qquad{} +\mbar^2 t \big[2-2L_m+2 L_t
    \non &\qquad{}\qquad{}\qquad{}\quad+C_{\mathrm{F}} \alpha_{\mathrm{s}}(\mu^2)
    (0.645693-0.30742 L_m-0.95493 L_m^2 \non
    &\qquad\qquad\qquad\qquad\qquad\qquad+2.69474 L_t+0.477465 L_m L_t+0.477465 L_t^2) \big] \non
  &\qquad{}\qquad{}+\mathcal{O}((\mbar^2 t)^2)
  \bigg\}+\mathcal{O}(\alpha_{\mathrm{s}}^2) \label{seriesasmall}
\end{align}
for $\mbar^2 t \ll 1$,
and
\begin{align}
r_a(\mbar)
&=\mbar t
\bigg\{
0.5+C_{\mathrm{F}} \alpha_{\mathrm{s}}(\mu^2)(-0.0554033+0.0596831 L_m+0.0596831 L_t) \non
&\qquad{}\qquad{}
+\frac{1}{\mbar^2 t} \left[0.25+C_{\mathrm{F}} \alpha_{\mathrm{s}}(\mu^2) (-0.0465134-0.0596831 L_m) \right] \non
&\qquad{}\qquad{}+\mathcal{O}((\mbar^2 t)^{-2})
\bigg\}+\mathcal{O}(\alpha_{\mathrm{s}}^2)
\end{align}
for $\mbar^2 t \gg 1$.  The perturbative series for $r_c(\mbar)$ can be
straightforwardly obtained from this expression. For $r_b$ the
perturbative series reads
\begin{align}
r_b(\mbar)
&=1-\mbar^2 t
\bigg\{
2+C_{\mathrm{F}} \alpha_{\mathrm{s}}(\mu^2) (0.589536+0.95493  L_t) \non
&\qquad{}\qquad{}
+\mbar^2 t \big[-4 L_m+4 L_t \non
&\qquad{}\qquad{}\qquad{}\qquad{}+C_{\mathrm{F}} \alpha_{\mathrm{s}}(\mu^2)
(-1.60014-1.41061 L_m-1.90986 L_m^2 \non
&\qquad{}\qquad{}\qquad{}\qquad{}\qquad{}\qquad{}\qquad{}
+3.32047 L_t+1.90986 L_t^2 )
\big] \non
&\qquad{}\qquad{}+\mathcal{O}((\mbar^2 t)^2)
\bigg\}+\mathcal{O}(\alpha_{\mathrm{s}}^2)  \label{seriescsmall}
\end{align}
for $\mbar^2 t \ll 1$,
and
\begin{align}
&r_b(\mbar) =\frac{1}{\mbar^2 t}
\bigg\{
1-C_{\mathrm{F}} \alpha_{\mathrm{s}}(\mu^2) (0.0753887+0.358099 L_m+0.119366 L_t) \non
&\qquad{}\qquad{}
+\frac{1}{\mbar^2 t} \big[-1.5
+C_{\mathrm{F}} \alpha_{\mathrm{s}}(\mu^2)
(0.3496+1.01461 L_m+0.417782 L_t)
\big] \non
&\qquad{}\qquad{}+\mathcal{O}((\mbar^2 t)^{-2})
\bigg\}+\mathcal{O}(\alpha_{\mathrm{s}}^2) \label{seriesclarge}
\end{align}
for $\mbar^2 t \gg 1$.  Higher orders in $\mbar^2 t$ and in
$1/(\mbar^2 t)$ can be easily obtained from the results presented in
app.~\ref{app:results}.  In addition,
the analytical expressions for these coefficients are available.
The above equations are presented only to illustrate the structure.

The dependence of these results on the renormalization scale $\mu$ is formally
of higher orders in perturbation theory. Nevertheless, in order to study their
numerical behavior, we need to choose a value for $\mu$. In the small-$t$
region, it is natural to set $\mu^2\sim 1/t$, since this drives
$\alpha_{\mathrm{s}}(\mu^2)$ into the perturbative region and also eliminates potentially
large logarithms. In the large-$t$ region, however, we need to prevent $\mu$
from entering the non-perturbative regime. Unless stated otherwise, our
central choice for $\mu$ which interpolates between these two requirements is
\begin{equation}
  \begin{aligned}
    \mu_{\rm int}(t)
    \equiv \sqrt{\mu_t^2+\mbar^2(\mbar)} \label{muint} ,
  \end{aligned}
\end{equation}
with
\begin{equation}
  \begin{aligned}
    \mu_t \equiv \frac{\rme^{-\EulerGamma/2}}{\sqrt{2t}} \label{mut}\,.
  \end{aligned}
\end{equation}
We do not claim that eq.~\eqref{muint} is an optimal choice for general
$1/\sqrt{t}$ and $\mbar$.  For the scope of this manuscript we prefer a simple prescription over a
dedicated study of the optimal renormalization scale.

As input parameters, we adopt~\cite{ParticleDataGroup:2024cfk}
\begin{equation}\label{eq:resu:kilt}
  \begin{aligned}
    \alpha_{\mathrm{s}}(M^2_{\mathrm{Z}})&=0.1180\,,\\
    \mbar_{\mathrm{s}}(2~{\rm GeV})&=93.5\,\text{MeV}\,,\\
    \mbar_{\mathrm{c}}(\mbar_{\mathrm{c}})&=1.2730\,\text{GeV}\,,\\
    \mbar_{\mathrm{b}}(\mbar_{\mathrm{b}})&=4.183\,\text{GeV}\,.
  \end{aligned}
\end{equation}
Using the five-loop beta function
\cite{Baikov:2016tgj,Herzog:2017ohr,Luthe:2017ttg} and the four-loop
decoupling relations for the coupling \cite{Chetyrkin:1997sg}, we obtain the
running coupling at our central scale choice. Similarly, the running mass at
the central scale is obtained based on the four-loop anomalous dimension of
the $\MSbar$ mass \cite{Vermaseren:1997fq}.

\begin{figure}
  \begin{center}
    \begin{tabular}{cc}
      \includegraphics[width=.45\textwidth]{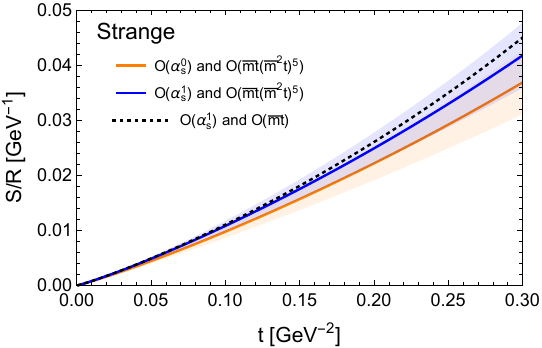} &
        \includegraphics[width=.45\textwidth]{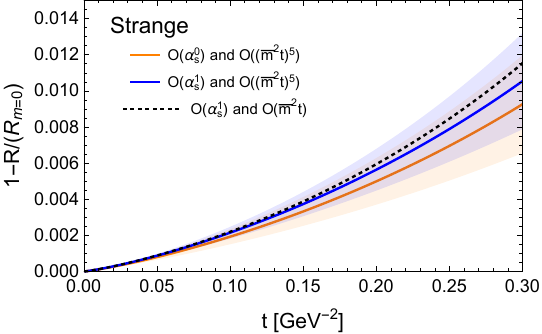}\\
      (a) & (b)\\
      \multicolumn{2}{c}{%
      \includegraphics[width=.45\textwidth]{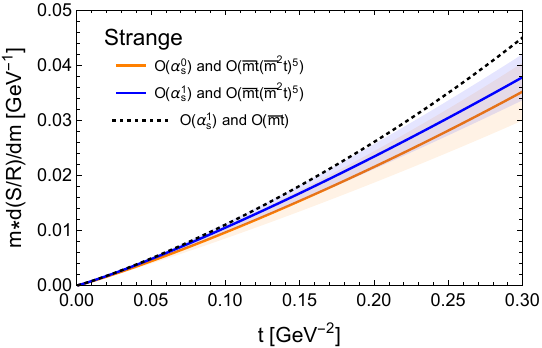}
      }\\
      \multicolumn{2}{c}{%
        (c)}
    \end{tabular}
    \caption{Perturbative results for the strange quark, with error bands
      estimated by varying the renormalization scale by a factor of two around
      its central value $\mu_t$, see eq.~\eqref{mut}.  The number of flavors
      is set to $n_{\mathrm{f}}=3$. The $\mathcal{O}(\alpha_{\mathrm{s}})$ result with the leading
      term in the $\mbar^2 t$-expansion is shown by the black dashed line for
      reference.}
\label{fig:strange}
  \end{center}
\end{figure}

The perturbative results of the observables defined in
eq.~\eqref{eq:resu:gulf} for the strange quark are shown in
fig.~\ref{fig:strange} where we subtract one from $r_b$ to suppress non-perturbative effects as discussed in sec.~\ref{sec:NP}.
Here we use the small-$\mbar^2 t$ expansion, as
$\mbar_s^2 t\lesssim 0.005$ in the range of these figures.  Therefore, rather than using
$\mu_\text{int}(t)$ of eq.~\eqref{muint}, we simply set the central
renormalization scale to $\mu=\mu_t$, defined by eq.~\eqref{mut}.  We
confirmed that the results agree with the numerical results retaining full
mass dependence at larger $t$ in the figures. The black dashed lines show the
leading term in the small-$\mbar^2 t$ expansion at $\mathcal{O}(\alpha_{\mathrm{s}})$.
The perturbative uncertainties are indicated by the bands whose width is
determined by
\begin{equation}\label{eq:resu:boil}
  \begin{aligned}
    \Delta &= \max_{\mu=\mu_t\,{\text{or}}\,2\mu_t} r - \min_{\mu=\mu_t\,{\text{or}}\,2\mu_t} r\,.
  \end{aligned}
\end{equation}
We do not
consider $\mu=\mu_t/2$ here, as it would extend into the non-perturbative
region.
For this variation of $\mu$, we use the one-loop mass anomalous dimension at $\mathcal{O}(\alpha_{\mathrm{s}}^0)$ and the one-loop beta-function and the two-loop mass anomalous dimension at $\mathcal{O}(\alpha_{\mathrm{s}})$.
The band is then
symmetrized around the central value for $r_a$, obtained by setting
$\mu=\mu_t$.


%
%

\begin{figure}
\begin{minipage}{0.5\hsize}
\begin{center}
    \includegraphics[width=7cm]{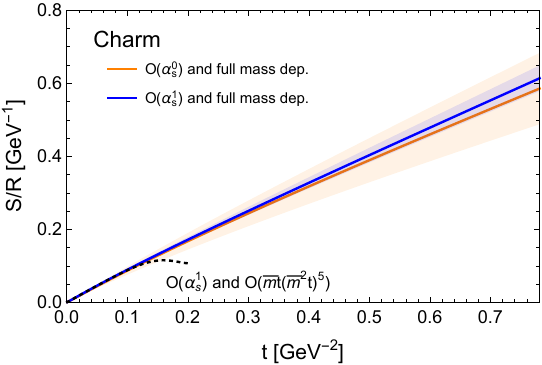}
\end{center}
\end{minipage}
\begin{minipage}{0.5\hsize}
\begin{center}
    \includegraphics[width=7cm]{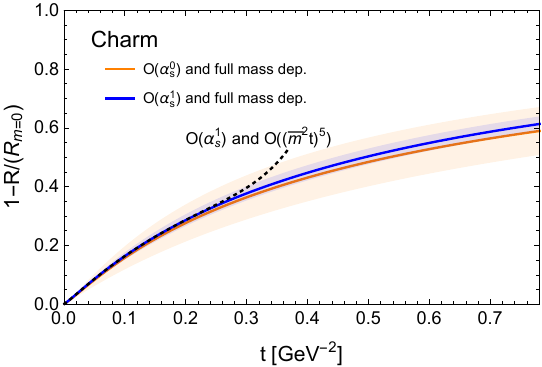}
\end{center}
\end{minipage}
\caption{ Perturbative results for the charm quark, with error bands estimated
  by varying the renormalization scale by a factor of two around its central
  value $\mu_{\rm int}(t)$, see eq.~\eqref{muint}. The number of flavors is
  set to $n_{\mathrm{f}}=4$.  The small-$\mbar^2 t$ expansion at $\mathcal{O}(\alpha_{\mathrm{s}})$
  is shown by the black dashed line for reference. The large-$\mbar^2 t$
  expansions have no relevance in this $t$-range.}
\label{fig:charm}
\end{figure}
\begin{figure}
\begin{minipage}{0.5\hsize}
\begin{center}
    \includegraphics[width=7cm]{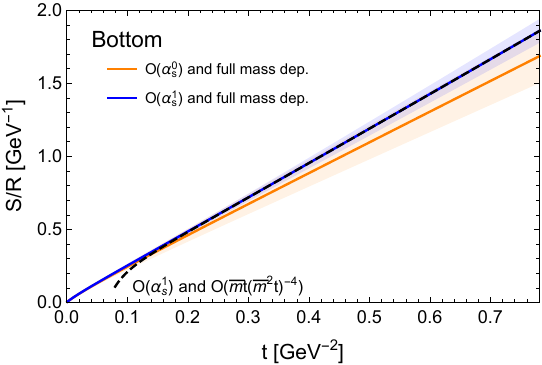}
\end{center}
\end{minipage}
\begin{minipage}{0.5\hsize}
\begin{center}
    \includegraphics[width=7cm]{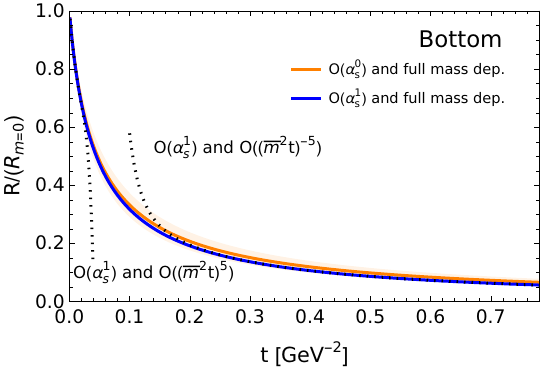}
\end{center}
\end{minipage}
\caption{ Perturbative results for the bottom quark, with error bands
  estimated by varying the renormalization scale by a factor of two around its
  central value $\mu_{\rm int}(t)$, see eq.~\eqref{muint}. The number of
  flavors is set to $n_{\mathrm{f}}=5$.
  The small- and large-$\mbar^2 t$ expansions at
  $\mathcal{O}(\alpha_{\mathrm{s}})$ are shown by the black dashed lines for
  reference. The small-$\mbar^2 t$ expansion for the left figure is not shown
  as it is valid only in a tiny region.}
\label{fig:bottom}
\end{figure}

The analogous results for the charm and bottom quark are shown in
fig.~\ref{fig:charm} and \ref{fig:bottom}, respectively.  Here
we use the numerical results retaining full mass dependence rather than the
expansions, since the validity of the latter does not extend to the full range
in $t$. Also, we only consider $r_a$ and $r_b$ in this case for reasons that
will be discussed in sec.~\ref{sec:NP}.\footnote{For the charm we also subtract the trivial one from $r_b$ for better visibility.}  As the central
renormalization scale, we choose $\mu=\mu_{\rm int}(t)$ in this case, defined
by eq.~\eqref{muint}. For reference, we remark that the two terms $\mu_t$ and
$\mbar$ contributing to $\mu_{\rm int}$ are equal at $t\approx
0.173$\,GeV$^{-2}$ for the charm quark, and at $t\approx 0.016$\,GeV$^{-2}$
for the bottom quark.  The perturbative uncertainties are estimated in analogy
to eq.~\eqref{eq:resu:boil}, but using $\mu=\mu_{\rm int}(t)/2$ and $2\mu_{\rm int}(t)$.

\begin{figure}[tbh]
\begin{minipage}{0.5\hsize}
\begin{center}
    \includegraphics[width=7cm]{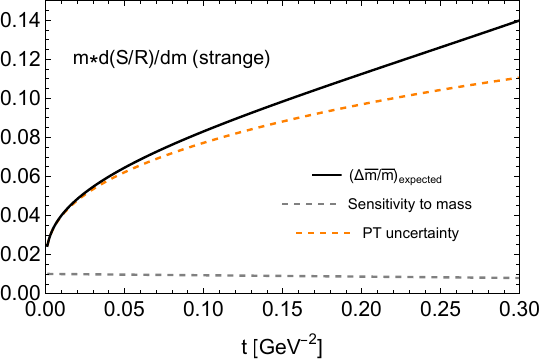}
\end{center}
\end{minipage}
\begin{minipage}{0.5\hsize}
\begin{center}
    \includegraphics[width=7cm]{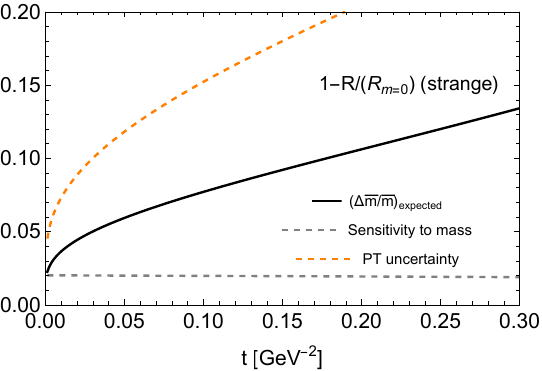}
\end{center}
\end{minipage}
\begin{minipage}{0.5\hsize}
\begin{center}
    \includegraphics[width=7cm]{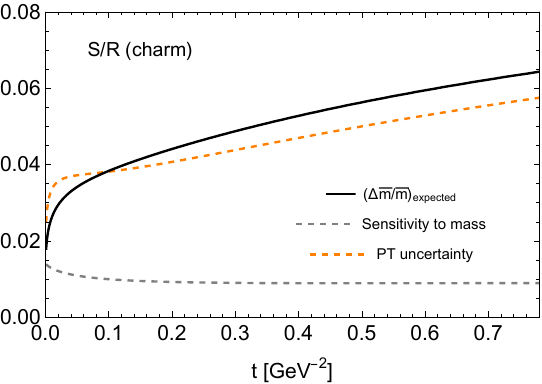}
\end{center}
\end{minipage}
\begin{minipage}{0.5\hsize}
\begin{center}
    \includegraphics[width=7cm]{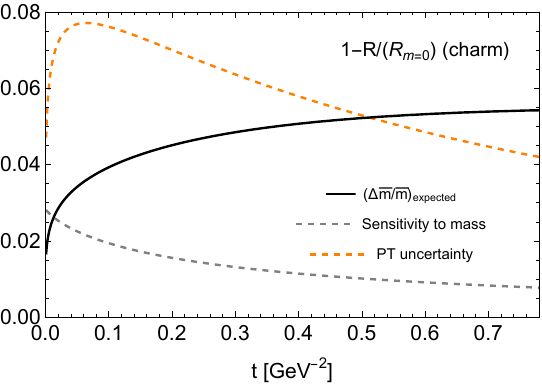}
\end{center}
\end{minipage}
\begin{minipage}{0.5\hsize}
\begin{center}
    \includegraphics[width=7cm]{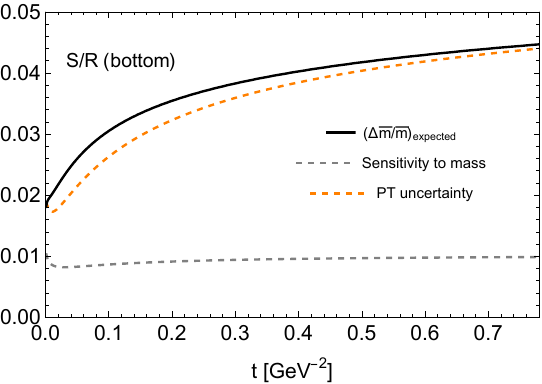}
\end{center}
\end{minipage}
\begin{minipage}{0.5\hsize}
\begin{center}
    \includegraphics[width=7cm]{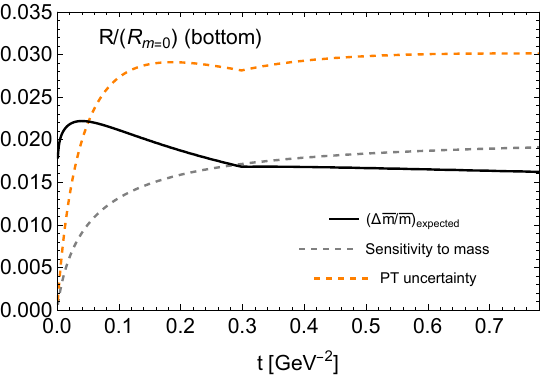}
\end{center}
\end{minipage}
\caption{Estimate of the expected accuracy on mass determination at the current
perturbative order $[\mathcal{O}(\alpha_{\mathrm{s}})]$.
See the text for details. }
\label{fig:EstimatedMassAccuracy}
\end{figure}

Let us now study the expected accuracy of a mass
determination using the three observables $r_a$, $r_b$, and $r_c$. To estimate
this accuracy, let us for the moment assume that an infinitely precise lattice
measurement of a quantity $r$ is available, denoted by $r_\text{exp}$, and
that the theoretical prediction $r(\mbar)$ of this quantity depends on the mass
$\mbar$. Neglecting also theoretical uncertainties, we could determine the
physical value $\mbar$ as the solution of
\begin{equation}\label{eq:resu:bene}
  \begin{aligned}
    r(\mbar) &= r_\text{exp}\,.
  \end{aligned}
\end{equation}
However, if the theoretical prediction carries an uncertainty $\delta r$, then
all one can say is that, to some level of confidence,
\begin{equation}\label{eq:resu:emil}
  \begin{aligned}
    r(\mbar) \in [r_\text{exp}-\delta r,r_\text{exp}+\delta r]\,.
  \end{aligned}
\end{equation}
Thus, we can determine $\mbar$ only up to $\pm\Delta \mbar$, where
\begin{equation}\label{eq:resu:agal}
  \begin{aligned}
    r(\mbar\pm \Delta \mbar)
    \approx r(\mbar) \pm r'(\mbar) \Delta \mbar
    \in [r_\text{exp}-\delta r,r_\text{exp}+\delta r]\,,
  \end{aligned}
\end{equation}
and therefore
\begin{equation}\label{eq:resu:kibe}
  \begin{aligned}
    \Delta \mbar &= \left|\frac{\delta r}{r'(\mbar)}\right|\,.
  \end{aligned}
\end{equation}
In fig.~\ref{fig:EstimatedMassAccuracy}, we study this quantity for the
observables $r_a$, $r_b$, and $r_c$. We only take into account the
perturbative uncertainty in $\delta r$ though; eventually, one would
have to take into account the uncertainty from the lattice input, of course.
The figures show the expected precision $\Delta \mbar/\mbar$ (solid black),  the relative
(perturbative) uncertainty $\delta r/r$
(\abbrev{PT} uncertainty),
and the dimensionless derivative $\mbar r'(\mbar)/r \times 0.01$
(where $0.01$ is multiplied for the sake of visibility in the figure) (Sensitivity to mass),
as a function of $t$, fixing $\mbar$ at the values given in
eq.~\eqref{eq:resu:kilt}.

In the case of the strange and charm quarks, where the small-$\mbar^2 t$ is
more relevant, we see the tendency that $\Delta \mbar/\mbar$ decreases towards smaller
values of $t$, because the perturbative uncertainty decreases in this
case. For the strange quark, the effect of the larger perturbative uncertainty
for $r_b$ compared to $r_c$ is partly compensated by the larger derivative,
leading to a similar expected accuracy on $\mbar$.  In the case of the
bottom quark, the figure on $r_b$ indicates the potential to extract its mass
value at $1 - 2$~\% already at this perturbative order.  We note that these
are crude estimates, based on the scale variation
around our central scale choices, eq.~\eqref{muint} or \eqref{mut}.
Therefore they are understood as order-of-magnitude estimates.

Besides the perturbative uncertainty due to missing higher orders, we also have to take into account the uncertainties from the lattice determinations for a full uncertainty estimate.
In addition, there are also non-perturbative corrections which we discuss in the next section.
We note, however, that the relevant  non-perturbative corrections for $\mbar^2 t \gg 1$ are currently unknown, as mentioned below,
and their systematic estimate remains an open problem.

\section{Discussion of non-perturbative effects}
\label{sec:NP}

So far we have neglected non-perturbative effects which will be particularly relevant for light quarks with $\mbar \ll \Lambda_{\rm QCD}$.
For $\mbar^2 t \ll 1$ we can estimate them based on the small-flow-time expansion discussed in sec.~\ref{sec:sftx}.
Taking the expressions from eqs.~\eqref{eq:S-sftx} and \eqref{eq:R-sftx}, we can split the observables into perturbative and non-perturbative contributions and find
\begin{equation}
  \bar{S} = \bar{S}_{\mathrm{PT}} + \zeta_\text{S}(t) \left[\langle\bar{\psi}\psi\rangle_{\mathrm{R}} - \langle\bar{\psi}\psi\rangle_{\mathrm{R,PT}} \right] + \mathcal{O}(t)
  \label{eq:S-non-pert}
\end{equation}
and
\begin{equation}
  \begin{split}
    \bar{R} = \bar{R}_{\mathrm{PT}} &+ \zeta_{21}(t) \frac{1}{g^2} \left[\langle F^a_{\mu\nu} F^a_{\mu\nu} \rangle_{\mathrm{R}} - \langle F^a_{\mu\nu} F^a_{\mu\nu} \rangle_{\mathrm{R,PT}} \right] \\
    &+ \zeta_{22}(t) \left[ \left\langle \bar{\psi} \overleftrightarrow{\slashed{D}} \psi \right\rangle_{\mathrm{R}} - \left\langle \bar{\psi} \overleftrightarrow{\slashed{D}} \psi \right\rangle_{\mathrm{R,PT}} \right] + \mathcal{O}(t) ,
  \end{split}
\end{equation}
where $\bar{S}_{\mathrm{PT}}$ and $\bar{R}_{\mathrm{PT}}$ represent the perturbative results calculated in this paper up to $\mathcal{O}(\alpha_{\mathrm{s}})$.
The differences $\left[\langle O \rangle_{\mathrm{R}} - \langle O \rangle_{\mathrm{R,PT}}\right]$ between the full, exact evaluation of the \abbrev{VEV} and its perturbative contributions,
which are non-zero for $m \neq 0$, single out the non-perturbative effects.

For $\bar{S}$ in eq.~\eqref{eq:S-non-pert}, the non-perturbative contribution, $\langle\bar{\psi}\psi\rangle_{\mathrm{R}} - \langle\bar{\psi}\psi\rangle_{\mathrm{R,PT}}$, is of $\mathcal{O}(\Lambda_{\rm QCD}^3)$ and can practically be more significant than the leading perturbative contribution $\bar{S}|_{\rm PT}=\mathcal{O}(\mbar/t)$. One can see this as follows.
For the strange quark
we are in the regime where $\mbar^2 \ll \Lambda_{\rm QCD}^2 \ll (8t)^{-1} \ll a^{-2}$ and
the ratio of the non-perturbative contribution to the perturbative contribution,
$\frac{\Lambda_{\rm QCD}^3}{\mbar/t}$, can exceed one due to
\be
\frac{\Lambda_{\rm QCD}}{\mbar} \frac{\Lambda_{\rm QCD}^2 a^2}{8} \ll \frac{\Lambda_{\rm QCD}^3}{\mbar/t}  \ll \frac{\Lambda_{\rm QCD}}{\mbar} \frac{1}{8} .
\label{NPsize}
\ee
Similarly, the leading finite mass effect in $\bar{R}|_{\rm PT}$, that is $\mathcal{O}(\mbar^2/t)$,
can be subdominant compared to the non-perturbative effects of $\mathcal{O}(\Lambda_{\rm QCD}^4)$.
This is not optimal for a mass determination.

Compared to the simple ratios $r_a$ and $r_b$ in eq.~\eqref{eq:resu:gulf}, observables like
\be
r_c(m) = m\frac{\dd}{\dd m}\left(\frac{S(t,m)}{R(t,m)}\right)
\qquad{}
\text{and}
\qquad{}
1 - r_b(m) = 1 - \frac{R(t,m)}{R(t,m=0)}
\ee
could be more suitable, because they suppress non-perturbative effects.
Assuming that the non-perturbative effects consist of
possible power terms of $m$ and $\Lambda_{\rm QCD}$,
and for instance, as for $\bar{\psi} \psi$,
\be
\langle\bar{\psi}\psi\rangle_{\mathrm{R}} - \langle\bar{\psi}\psi\rangle_{\mathrm{R,PT}}
=c_0 \Lambda_{\rm QCD}^3+c_1 m \Lambda_{\rm QCD}^2+c_2 m^2 \Lambda_{\rm QCD} ,
\ee
where $c_{0,1,2}$ are constants,
one can see that the largest non-perturbative
contribution, $c_0 \Lambda_{\rm QCD}^3$, which is independent of $m$,
disappears due to the derivative in $r_c$.
Hence,
the leading non-perturbative effect
reduces to $\mathcal{O}(m \Lambda_{\rm QCD}^2)$, whose
magnitude relative to the perturbative effect $m/t$
reduces to $\mathcal{O}(t \Lambda_{\rm QCD}^2)$,
which is much smaller than one and can be well controlled.
As for $1-R/(R|_{m=0})$,
the leading non-perturbative effect is
$\mathcal{O}(m \Lambda_{\rm QCD}^3 t^2)$,
which is smaller than that found in $R/(R|_{m=0})$,
but still can be large compared with the perturbative effect $m^2 t$.
Based on these observations,
we conclude that $r_c$ is the most suitable quantity for light quark mass determination among $r_{a,b,c}$ and $1-r_b$.

For heavy quarks,
the non-perturbative effects are properly suppressed
already in $r_a$ and $r_b$, as long as $\mbar^2 t \ll 1$,
by applying a similar argument as above for $\Lambda_{\rm QCD}^2 \ll \mbar^2\ll (8t)^{-1} \ll a^{-2}$.
This is why $r_c$ is studied only for the strange quark.

For $\mbar^2 t \gg 1$ we know little about non-perturbative effects.
This study would be worth pursuing in the future.


\section{Conclusions}
\label{sec:conclusions}

We proposed a method to determine quark masses using the gradient flow
by considering ratios of \abbrev{VEV}s of flowed quark-bilinear operators. They are
scheme and gauge invariant and can be calculated both on the lattice and in
perturbation theory.  We evaluated these \abbrev{VEV}s perturbatively up to two loops,
i.e.\ $\mathcal{O}(\alpha_{\mathrm{s}})$, including finite mass effects.  Compared to the
already known perturbative results in the small-$\mbar^2t$ limit, we found
that the higher-order terms of the small-$\mbar^2t$ expansion, the
large-$\mbar^2t$ expansion, and full mass dependence are crucial for studying
particularly charm and bottom quarks. The \abbrev{VEV}
ratios exhibit good perturbative behavior.  We also estimated the expected
accuracy achieved in the mass determination from the perturbative viewpoint.

We presented the small- and large-$\mbar^2 t$ expansions of these \abbrev{VEV}s, as
well as fully numerical perturbative series retaining the full mass
dependence.  To calculate the expansions, we developed a method based on the
Laplace transform rather than the often employed technique called ``strategy
of regions.''  The Laplace transform with respect to a combination of the
external scales $m$ and $t$ is considered, whose singularities and residues
give expansions of loop integrals.  Although the Laplace transforms have been
used in somewhat analogous manners in the literature, to our understanding,
this is the first application to expanding  integrals beyond one loop in terms of
external scales.

In order to improve the accuracy of the quark mass determination and to be competitive with existing
methods, it may be required to compute higher orders in the perturbative
series. Even though the technology for such calculations is available (see,
e.g.~refs.~\cite{Artz:2019bpr,Harlander:2024vmn}), they are beyond the scope of
the current paper.

Finally, the method proposed in this paper requires lattice results of the
flowed quark bilinear operators for determining the quark mass.  These
quantities have been simulated in
refs.~\cite{Taniguchi:2016ofw,Taniguchi:2020mgg} for light quarks ($u, d, s$).
It will be interesting to study which accuracy can be achieved already with
these simulations following the proposed method and what is required for a
competitive extraction.

\acknowledgments We thank Robert Mason for checking parts of our numerical
results.  H.T. thanks Yukinari Sumino, Hiroshi Suzuki, and especially Masakiyo
Kitazawa for helpful discussions. The work of
H.T. was supported by JSPS KAKENHI Grant Numbers JP19K14711 and
JP23K13110. H.T. is the Yukawa Research Fellow supported by Yukawa Memorial
Foundation. The work of R.V.H. was supported by the Deutsche
Forschungsgemeinschaft (DFG, German Research Foundation) under grant 396021762
- TRR~257.
The work of F.L.~was supported by the Swiss National Science Foundation (SNSF) under contract \href{https://data.snf.ch/grants/grant/211209}{TMSGI2\_211209}.


\appendix


\section{List of loop integrals}
\label{sec:ints}

In this appendix we list all of the integrals appearing in our calculation of $S$ or $R$ described in sec.~\ref{sec:calc}.
Some of the following integrals do not contribute directly, but are introduced to facilitate the evaluation of other integrals.

The scalar integrals relevant for $S$ are
\begin{align}
&I_1=\int_0^t \dd s \, \int_{p,k} \frac{m}{k^2 (p^2+m^2)} \rme^{-2tp^2-2sk^2} , \non
&I_2=\int_0^t \dd s \int_0^s \dd s' \int_{p,k} \frac{m p^2}{k^2 (p^2+m^2)} \rme^{-(2t-s+s') p^2-(s+s') k^2-(s-s') (k-p)^2} ,\non
&I_3=\int_0^t \dd s \int_{p,k} \frac{m}{k^2 (p^2+m^2)} \rme^{-(2t-s) p^2-s k^2-s(k-p)^2} , \non
&I_4=\int_0^t \dd s \int_0^t \dd s' \int_{p,k} \frac{m (p-k)^2}{k^2((p-k)^2+m^2)} \rme^{-(2t-s-s')p^2-(s+s') k^2-(s+s') (k-p)^2} ,\non
&I_5=\int_{p,k} \frac{m}{k^2(p^2+m^2)((p-k)^2+m^2)}\rme^{-tp^2-tk^2-t(k-p)^2} ,\non
&I_6=\int_0^t \dd s \int_{p,k} \frac{m}{k^2 ((p-k)^2+m^2)} \rme^{-(2t-s)p^2-sk^2-s(k-p)^2} ,\non
&I_7=\int_0^t \dd s \int_{p,k} \frac{m^3}{k^2(p^2+m^2)((p-k)^2+m^2)} \rme^{-(2t-s)p^2-sk^2-s(k-p)^2} ,\non
&I_8=\int_{p,k} \frac{m}{k^2(p^2+m^2)((p-k)^2+m^2)}\rme^{-2tp^2} ,\non
&I_9=\int_{p,k} \frac{m}{k^2(p^2+m^2)^2}\rme^{-2tp^2}  ,\non
&I_{10}=\int_{p,k} \frac{m}{(p^2+m^2)^2((p-k)^2+m^2)}\rme^{-2tp^2}  ,\non
&I_{11}=\int_{p,k} \frac{m^3}{k^2(p^2+m^2)^2((p-k)^2+m^2)}\rme^{-2tp^2}
\end{align}
and the scalar integrals relevant for $R$ are
\begin{align}
J_1&=\int_0^t \dd s \, \int_{p,k}  \frac{ (p-k)^2}{k^2 ((p-k)^2+m^2)}  \rme^{-(t-s) p^2-(t+s) k^2-(t+s)(p-k)^2}  \non
J_2&=\int_{p,k} \frac{1}{(p^2+m^2)((p-k)^2+m^2)} \rme^{-t p^2-t k^2-t (p-k)^2} \non
J_3&=\int_{p,k} \frac{1}{k^2 (p^2+m^2)} \rme^{-t p^2-t k^2-t (p-k)^2}  \non
J_4&=\int_{p,k} \frac{m^2}{k^2 (p^2+m^2) ((p-k)^2+m^2)} \rme^{-t p^2-t k^2-t (p-k)^2} \non
J_5&=\int_0^t \dd s \, \int_{p,k}  \frac{p^2}{k^2 (p^2+m^2)} \rme^{-2 t p^2-2s k^2}  \non
J_6&=\int_0^t \dd s \int_0^s \dd s' \,  \int_{p,k}   \frac{(p^2)^2}{k^2 (p^2+m^2)} \rme^{-(2 t-s+s') p^2-(s+s')k^2-(s-s') (k-p)^2} \non
J_7&=\int_0^t \dd s \, \int_{p,k}  \frac{p^2}{k^2 (p^2+m^2)} \rme^{-(2 t-s) p^2-sk^2-s (k-p)^2} \non
J_8&=\int_0^t \dd s \int_0^t \dd s' \, \int_{p,k}  \frac{((p-k)^2)^2}{k^2((p-k)^2+m^2)} \rme^{-(2t-s-s')p^2-(s+s') k^2-(s+s')(k-p)^2} \non
J_{9}&=\int_0^t \dd s \, \int_{p,k}  \frac{(p-k)^2}{k^2((p-k)^2+m^2)} \rme^{-(2t-s)p^2-s k^2-s(k-p)^2} \non
J_{10}&=\int_0^t \dd s \, \int_{p,k} \frac{1}{k^2} \rme^{-(2t-s)p^2-s k^2-s(k-p)^2} \non
J_{11}&=\int_0^t \dd s \, \int_{p,k}  \frac{m^2}{k^2 (p^2+m^2)} \rme^{-(2t-s)p^2-s k^2-s(k-p)^2} \non
J_{12}&=\int_0^t \dd s \, \int_{p,k}  \frac{m^2}{k^2 ((p-k)^2+m^2)} \rme^{-(2t-s)p^2-s k^2-s(k-p)^2} \non
J_{13}&=\int_0^t \dd s \, \int_{p,k} \frac{m^4}{k^2 (p^2+m^2) ((p-k)^2+m^2)} \rme^{-(2t-s)p^2-s k^2-s(k-p)^2} \non
J_{14}&=\int_{p,k} \frac{m^2}{k^2 (p^2+m^2) ((p-k)^2+m^2)} \rme^{-2 t p^2} \non
J_{15}&=\int_{p,k} \frac{1}{(p^2+m^2) ((p-k)^2+m^2)} \rme^{-2tp^2} \non
J_{16}&=\int_{p,k} \frac{1}{k^2 (p^2+m^2)} \rme^{-2tp^2} \non
J_{17}&=\int_{p,k} \frac{1}{k^2 ((p-k)^2+m^2)} \rme^{-2tp^2} \non
J_{18}&=\int_{p,k} \frac{m^2}{(p^2+m^2)^2 ((p-k)^2+m^2)} \rme^{-2tp^2}  \non
J_{19}&=\int_{p,k} \frac{m^2}{k^2 ((p^2+m^2)^2)^2} \rme^{-2tp^2} \non
J_{20}&=\int_{p,k} \frac{m^4}{k^2 (p^2+m^2)^2 ((p-k)^2+m^2)} \rme^{-2 t p^2} .
\end{align}
In the above expressions, $m$ denotes the bare mass and
$k$ denotes the gluon loop momentum.


\section{Choosing $v_0$ for the integral $J_{17}$}\label{sec:J17}

In the Laplace transform of $J_{17}$, we find singularities at $v=-\ep$ and
$v=0$.  Both of them give $\mathcal{O}(1/t^2)$ terms upon the $\ep$-expansion,
because a singularity at $v$ corresponds to a term of the form
$m^{2v} t^{-2+v+2 \ep}$.  We
need to clarify whether these singularities contribute to either the small- or
large-$m^2 t$ expansion.  (This issue is related to which value $v_0$ should
be set to.  In principle, one can determine this by searching for $v=v_0$
where every step of the calculation is valid, but this is not always
practical.)

First, it turns out that the $v=0$ singularity
contributes to the small-$m^2 t$ expansion.
This is because the residue at $v=0$ is identical
to the original loop integral with the mass in the propagator sent to zero.
Secondly, as for $v=-\ep$, it is useful to consider
a bound on $J_{17}$ in $d$ dimensions,
\begin{align}
|J_{17}|
& < \int_{p,k} \frac{1}{k^2 (p-k)^2} \rme^{-2 t p^2} \non
&=C \int_{p} \frac{1}{(p^2)^{(4-d)/2}}  \rme^{-2 t p^2} \non
&=C' t^{2-d}=C' t^{-2+2 \ep} . \label{bound}
\end{align}
In the first inequality, we used $1/[(p-k)^2+m^2] < 1/(p-k)^2$.
The bound~\eqref{bound} is valid as long as $2<d<4$;
the $k$-integral is \abbrev{UV} divergent for $d \geq 4$
and infrared divergent for $d \leq 2$.
Therefore, we assume this range of $d$ and thus $\ep>0$.
$C$ and $C'$ are positive constants.
From this bound, we can see that
the $v=-\ep$ singularity, which gives $\mathcal{O}(m^{-2 \ep} t^{-2+\ep})$,
contributes to the large-$m^2 t$ expansion;
if it contributed to
the small-$m^2 t$ expansion,
it would exceed the bound for small $t$.
To summarize, we should choose $-\ep < v_0 < 0$
in this calculation.

In the calculation of $J_{17}$, using eq.~\eqref{formulatwo}
after the change of momenta $p \to p+k$ and then $k \to -k$, leads to a singular expression.
One can circumvent this by introducing
``the $s$-parameter in eq.~\eqref{formulatwo}''
as a regulator and then sending it to zero.



\section{Formulae}
\label{sec:C}
We collect formulae concerning the Laplace transform
and loop integrals.

\subsection{Laplace transform}

\begin{align}
&\int_0^{\infty} \dd z \, z^{-v-1/2} \frac{1}{P^2+z}
=\frac{\pi}{\cos(\pi v)} \frac{1}{(P^2)^{v+1/2}} , \non
&\int_0^{\infty} \dd z \, z^{-v-1/2} \frac{1}{(P^2+z)^2}
= \frac{1}{2} (1+2v)\frac{\pi}{\cos(\pi v)} \frac{1}{(P^2)^{v+3/2}} . \label{Laplace}
\end{align}
$P$ denotes a linear combination of (dimensionless) loop momenta.

\subsection{Loop integrals}

\begin{align}
&\int_p \frac{1}{(p^2)^a} \rme^{-s p^2}=\frac{1}{(4 \pi)^{d/2}} \frac{\Gamma(d/2-a)}{\Gamma(d/2)} s^{a-d/2} \quad{} \mbox{\cite{Luscher:2010iy}} \label{formulazero}  , \\
&\int_{p,k} \frac{\rme^{-s p^2-u k^2-v(k-p)^2}}{(k^2)^a}=\frac{1}{(4 \pi)^d} \frac{\Gamma(d/2-a)}{\Gamma(d/2)} \frac{1}{(s+v)^a} \frac{1}{(uv+su+sv)^{d/2-a}} , \label{formulaone} \\
& \int_{p,k} \frac{1}{(k^2)^a (p^2)^b} \rme^{-s p^2-u k^2-v (k-p)^2} \non
&=\frac{1}{(4 \pi)^d} \frac{\Gamma(d/2-b)}{\Gamma(d/2)\Gamma(a)} \frac{1}{(u+v)^b} \frac{1}{(uv+su+sv)^{d/2-b}} \non
&\quad{} \times \bigg[
\left( \frac{uv+su+sv}{s+v} \right)^a \frac{\Gamma(a) \Gamma(d/2-a-b)}{\Gamma(d/2-b)}  {}_2F_1(a;b;1+a+b-d/2;\frac{uv+su+sv}{(s+v)(u+v)}) \non
&\qquad{}\quad{} +\frac{1}{\Gamma(b)} (u+v)^a \left(\frac{(s+v)(u+v)}{uv+su+sv} \right)^{b-d/2} \Gamma(a+b-d/2) \Gamma(d/2-a) \non
&\qquad{}\qquad{} \times {}_2F_1(\frac{d}{2}-a,\frac{d}{2}-b;\frac{d}{2}+1-a-b; \frac{uv+su+sv}{(s+v)(u+v)}) \bigg] , \label{formulatwo}\\
&\int_{k,p} \frac{1}{(k^2)^a [xp^2+(1-x) (k-p)^2]^b} \rme^{-sp^2-uk^2-v(k-p)^2}   \non
&=\frac{1}{(4\pi)^d}
\frac{\Gamma(d/2-a)}{\Gamma(b)\Gamma(d/2)} \non
&\quad{} \times \int_0^{\infty} \dd \beta \beta^{b-1}
\frac{1}{(s+v+\beta)^a} \frac{1}{[u(v+\beta(1-x))+(s+\beta x)u+(s+\beta x) (v+\beta(1-x))]^{d/2-a}}  . \label{formulatwoFeynman}
\end{align}
In the last equation, $x$ corresponds to a Fyenmal integral parameter ($0 \leq x \leq 1$).
${}_2F_1(a,b;c;z)$ denotes the hypergeometric function, which is defined by
\be
{}_2F_1(a,b;c;z) \equiv \sum_{n=0}^{\infty} \frac{(a)_n (b)_n}{(c)_n} \frac{z^n}{n!}
\ee
for $|z|<1$ and a non-positive integer $c$. $(...)_n$ is the Pochhammer symbol $(a)_n=\Gamma(a+n)/\Gamma(a)$.
In our calculation, we often use a formula to
change the $z$ variable as $z \to 1-z$,
\begin{align}
{}_2F_1(a,b;c;z)
&=\frac{\Gamma(c)\Gamma(c-a-b)}{\Gamma(c-a) \Gamma(c-b)} {}_2F_1(a,b;a+b+1-c;1-z) \non
&\quad{}+\frac{\Gamma(c) \Gamma(a+b-c)}{\Gamma(a) \Gamma(b)} (1-z)^{c-a-b}  {}_2F_1(c-a,c-b;1+c-a-b;1-z) \label{rewritehyp}
\end{align}
when the alternative expression is found more convenient
to understand the structure of singularities.


\section{Perturbative series}\label{app:results}

We present the two-loop expressions in the small- and large-$\mbar^2 t$ limits.
$\overline{S}$ for small-$\mbar^2 t$ is given by
\be
\overline{S}
=- \frac{N_{\mathrm{c}}}{8 \pi^2} \frac{\mbar}{t} \sum_{n,k=0}^{\infty} C^{S,  \ll 1}_{n,k}(\mbar^2, t, \mu^2) (\mbar^2 t)^k \lt(\frac{\alpha_{\mathrm{s}}(\mu^2)}{\pi} \rt)^n
\ee
where
\begin{align}
C^{S,  \ll 1}_{0,0} &=1 , \non
C^{S,  \ll 1}_{0,1} &=2 (L_t-L_m) , \non
C^{S,  \ll 1}_{0,2} &= 4 (L_t-L_m-1) , \non
C^{S,  \ll 1}_{0,3} &= 4 L_t-4 L_m-6 , \non
C^{S,  \ll 1}_{0,4} &= \frac{4}{9} \left(6 L_t-6 L_m-11\right) , \non
C^{S,  \ll 1}_{0,5} &=\frac{1}{9} \left(12 L_t -12L_m-25\right) ,
\end{align}
and
\begin{align}
C^{S,  \ll 1}_{1,0}
&= C_{\mathrm{F}} (1+2 \log 2) , \nonumber \\
C^{S,  \ll 1}_{1,1}
&= C_{\mathrm{F}} \bigg(3+10 \log 2-9 \log 3 - 6 \, \text{Li}_2\left(\frac{1}{4}\right) - 4 L_m    - 3 L_m^2 + 7 L_t + 3 L_m L_t \bigg) , \nonumber \\
C^{S,  \ll 1}_{1,2}
&= C_{\mathrm{F}} \bigg(-11.5053-15.25 L_m-9 L_m^2+9.25 L_t+6 L_m L_t+3 L_t^2 \bigg) , \nonumber \\
C^{S,  \ll 1}_{1,3}
&=  C_{\mathrm{F}} \bigg(-12.6989-20.4375 L_m-11 L_m^2 -0.5625 L_t+4 L_m L_t+7 L_t^2\bigg) , \nonumber \\
C^{S,  \ll 1}_{1,4}
&= C_{\mathrm{F}} \bigg(-4.2438-13.3656 L_m-7.5 L_m^2
-11.9677 L_t-L_m L_t+8.5 L_t^2
\bigg) , \nonumber \\
C^{S,  \ll 1}_{1,5}
&= C_{\mathrm{F}} \bigg(
4.2571-2.62156 L_m-2.75 L_m^2
-16.2118 L_t-4.5 L_m L_t+7.25 L_t^2
\bigg)  .
\end{align}
The log independent coefficients are valid
to the numbers of significant digits shown.
On the other hand, the coefficients of logarithmic terms
are calculated to higher precision
and numerical values such as $4.000\cdots$ are
replaced by the integer numbers,
and also, for instance, $7.5000\cdots$
of the $L_m^2$ coefficient in
$C_{1,4}^{S, \ll 1}$ is shown as $7.5$.

For large-$\mbar^2 t$, the expanion reads
\be
\overline{S}
=-\frac{N_{\mathrm{c}}}{16 \pi^2} \frac{1}{\mbar t^2} \sum_{n,k=0}^{\infty} C^{S,  \gg 1}_{n,k}(\mbar^2, t, \mu^2) (\mbar^2 t)^{-k} \left( \frac{\alpha_{\mathrm{s}}(\mu^2)}{\pi} \right)^n
\ee
where
\begin{align}
C^{S,  \gg 1}_{0,0} &= 1, \nonumber \\
C^{S,  \gg 1}_{0,1} &= -1, \nonumber \\
C^{S,  \gg 1}_{0,2} &= \frac{3}{2}, \nonumber \\
C^{S,  \gg 1}_{0,3} &= -3, \nonumber \\
C^{S,  \gg 1}_{0,4} &= \frac{15}{2}, \nonumber \\
C^{S,  \gg 1}_{0,5} &= -\frac{45}{2}, 
\end{align}
and
\begin{align}
C^{S,  \gg 1}_{1,0} &= -\frac{C_{\mathrm{F}}}{4} \bigg(2 + 3 L_m+3 L_t+6 \log{2}-9 \log{3}\bigg), \nonumber \\
C^{S,  \gg 1}_{1,1} &= \frac{C_{\mathrm{F}}}{16} \bigg(22+27 L_m+21 L_t+42 \log{2}-51 \log{3}\bigg), \nonumber \\
C^{S,  \gg 1}_{1,2} &= -\frac{3 C_{\mathrm{F}}}{64} \bigg(68+87 L_m+57 L_t+114 \log{2}-127\log{3}\bigg), \nonumber \\
C^{S,  \gg 1}_{1,3} &= \frac{C_{\mathrm{F}}}{768}
\bigg(6634+8775L_m+5049 L_t+10098 \log{2}-10719 \log{3}\bigg), \nonumber \\
C^{S,  \gg 1}_{1,4} &= -\frac{C_{\mathrm{F}}}{9216}
\bigg(253556 +341415 L_m+176985 L_t+353970\log{2} - 364095\log{3} \bigg), \nonumber \\
C^{S,  \gg 1}_{1,5} &= \frac{5 C_{\mathrm{F}}}{36864}
\bigg(746314+1012095 L_m+480897 L_t+961794 \log{2}-967383 \log{3} \bigg). 
\end{align}

$\overline{R}$ for small-$\mbar^2 t$ is given by
\be
\overline{R}
=-\frac{N_{\mathrm{c}}}{8 \pi^2} \frac{1}{t^2} \sum_{n,k=0}^{\infty} C^{R,  \ll 1}_{n,k}(\mbar^2, t, \mu^2) (\mbar^2 t)^k \lt( \frac{\alpha_{\mathrm{s}}(\mu^2)}{\pi} \rt)^n
\ee
where
\begin{align}
C^{R,  \ll 1}_{0,0} &= 1, \nonumber \\
C^{R,  \ll 1}_{0,1} &= -2, \nonumber \\
C^{R,  \ll 1}_{0,2} &= 4(L_m-L_t), \nonumber \\
C^{R,  \ll 1}_{0,3} &= 8(1 + L_m- L_t ), \nonumber \\
C^{R,  \ll 1}_{0,4} &= 12 +8 L_m-8 L_t , \nonumber \\
C^{R,  \ll 1}_{0,5} &= \frac{8}{9} \left(11 + 6 L_m- 6 L_t \right).
\end{align}
and
\begin{align}
C^{R,  \ll 1}_{1,0} &= \frac{C_{\mathrm{F}}}{4} \left(3 \log{3} + 4 \log{2} - 3 L_t\right), \nonumber \\
C^{R,  \ll 1}_{1,1} &= -\frac{C_{\mathrm{F}}}{2} \left(7 + 4 \log{2} + 3 L_t\right), \nonumber \\
C^{R,  \ll 1}_{1,2} &= -\frac{C_{\mathrm{F}}}{4} \bigg(25+120 \log{2}-108 \log{3} -36 \, \text{Li}_2\left(\frac{1}{4}\right) - 42 L_m
 - 24 L_m^2 + 66 L_t + 12 L_t L_m + 12 L_t^2 \bigg), \nonumber \\
C^{R,  \ll 1}_{1,3}
&= C_{\mathrm{F}}
(26.9641+29.8333 L_m+14 L_m^2-11.8333 L_t+2 L_m L_t-16 L_t^2), \nonumber \\
C^{R,  \ll 1}_{1,4}
&= C_{\mathrm{F}} (
22.0263+22.1667L_m+9.5 L_m^2+28.8333 L_t+23 L_m L_t-32.5 L_t^2
), \nonumber \\
C^{R,  \ll 1}_{1,5}
&=C_{\mathrm{F}} (
-11.3406-17.6604 L_m-6 L_m^2+75.6604 L_t+48 L_m L_t-42 L_t^2
).
\end{align}

For large-$\mbar^2 t$, it is given by
\be
\overline{R}
=-\frac{N_{\mathrm{c}}}{8 \pi^2} \frac{1}{\mbar^2 t^3} \sum_{n,k=0}^{\infty} C^{R,  \gg 1}_{n,k}(\mbar^2, t, \mu^2) (\mbar^2 t)^{-k} \lt( \frac{\alpha_{\mathrm{s}} (\mu^2)}{\pi} \rt)^n
\ee
where
\begin{align}
C^{R,  \gg 1}_{0,0} &= 1, \nonumber \\
C^{R,  \gg 1}_{0,1} &=-\frac{3}{2} , \nonumber \\
C^{R,  \gg 1}_{0,2} &= 3 , \nonumber \\
C^{R,  \gg 1}_{0,3} &= -\frac{15}{2}, \nonumber \\
C^{R,  \gg 1}_{0,4} &= \frac{45}{2},
\end{align}
and
\begin{align}
C^{R,  \gg 1}_{1,0} &= -\frac{C_{\mathrm{F}}}{16}
\bigg(4+18 L_m+18 L_t+36 \log{2}-45 \log{3}  \bigg), \nonumber \\
C^{R,  \gg 1}_{1,1}
&=\frac{C_{\mathrm{F}}}{64}
\bigg( 94 + 204 L_m+156L_t+312 \log{2}-351 \log{3}  \bigg), \nonumber \\
C^{R,  \gg 1}_{1,2}
&= -\frac{C_{\mathrm{F}}}{2304}
\bigg(
11686+22086 L_m+14202 L_t+28404 \log{2}-30267 \log{3}
\bigg), \nonumber \\
C^{R,  \gg 1}_{1,3}
&= \frac{C_{\mathrm{F}}}{9216}
\bigg(
168494+298080L_m+168480 L_t+336 960 \log{2}-347085 \log{3}
\bigg), \nonumber \\
C^{R,  \gg 1}_{1,4} &=-\frac{C_{\mathrm{F}}}{12288}
\bigg(
891478+1510650 L_m+770310 L_t+1540620 \log{2}-1549935 \log{3}
\bigg).
\end{align}
As mentioned, the analytic expressions of
$C^{S, \ll 1}_{1,0}$, $C^{S, \ll 1}_{1,1}$, $C^{R, \ll 1}_{1,0}$, $C^{R, \ll 1}_{1,1}$,
and $C^{R, \ll 1}_{1,2}$ are obtained by the method explained in sec.~\ref{sec:sftx}.

We provide these expansions as ancillary files that can be loaded in \texttt{Mathematica}.

Based on the numerical computation explained in sec.~\ref{sec:numerical}
we provide a grid for dimensionless $\bar{S}$ and $\bar{R}$.
The format is
$\{\mbar^2(\mu) t, \frac{t}{\mbar(\mu)} \bar{S} \}$ or
$\{\mbar^2(\mu) t, t^2 \bar{R} \}$,
where the perturbative series contain $\alpha_s(\mu^2)$, $\log(t \mu^2)$,
and the color factors as parameters.
Also the uncertainties of the perturbative coefficients are labeled by $\delta s[...]$
or $\delta r[...]$. Setting them to $0$ ($\pm 1$) corresponds to
the central values (estimated upper/lower bounds) of the perturbative coefficients.


\bibliographystyle{JHEP}
\bibliography{GradientFlow} 

\end{document}